\documentclass[aps,pre,reprint,superscriptaddress,amsmath,amssymb,showpacs,showkeys]{revtex4-1}

\usepackage{graphicx}
\usepackage{dcolumn}
\usepackage{bm}

\usepackage{mathrsfs}
\usepackage{txfonts}
\usepackage{balance} \balance 
\usepackage{array}
\usepackage{caption2}
\usepackage{subfigure}
\usepackage{float}

\begin{document}

\preprint{APS/123-QED}


\title{Koopman analysis in oscillator synchronization}

\author{Jing Hu}
\affiliation{School of Science, Beijing University of Posts and Telecommunications, Beijing 100876, China}

\author{Yueheng Lan}
\email{Email address : lanyh@bupt.edu.cn}
\affiliation{School of Science, Beijing University of Posts and Telecommunications, Beijing 100876, China}
\affiliation{State Key Lab of Information Photonics and Optical Communications, Beijing University of Posts and Telecommunications, Beijing 100876, China}

\date{\today}

\begin{abstract}
Synchronization is an important dynamical phenomenon in coupled nonlinear systems, which has been studied extensively in recent years. However, analysis focused on individual orbits seems hard to extend to complex systems while a global statistical approach is overly cursory. Koopman operator technique seems to well balance the two approaches. In this paper, we extend Koopman analysis to the study of synchronization of coupled oscillators by extracting important eigenvalues and eigenfunctions from the observed time series. A renormalization group analysis is designed to derive an analytic approximation of the eigenfunction in case of weak coupling that dominates the oscillation. For moderate or strong couplings, numerical computation further confirms the importance of the average frequencies and the associated eigenfunctions. The synchronization transition points could be located with quite high accuracy by checking the correlation of neighbouring eigenfunctions at different coupling strengths, which is readily applied to other nonlinear systems.

\end{abstract}

\maketitle


\section{Introduction}
With the rapid development of science and technology, huge amount of data is generated and collected each day, concerning with complex systems involving natural phenomena, engineering operations or human activities. Nevertheless, it is nearly impossible to get all the details of a large system and measurement could only be made for a small number of observables which are functions of the system state. In the theory of low-dimensional nonlinear dynamics, the state space reconstruction is possible even for a scalar observation~\cite{brunton2017chaos} since information is highly mixed when the dynamics is chaotic. However, in high-dimensional systems, the observations are so scant that much information is simply missed from the data. On the other hand, measurement is often contaminated with various noise that could be very disturbing. How to extract important information about the system from these partial observations remains a major challenge in both theory and practice.

At this point, the traditional focus on a single orbit may not be so appealing and it should be more fruitful to view the evolution of all observables as a whole, which may suppress possible noise and excavate stable information simultaneously from all the components as well as from their temporal correlation.
The evolution of an observable as a function in the state space is described by the Koopman operator proposed by Bernard Koopman and John von Neumann in 1930 ~\cite{koopman1931hamiltonian,koopman1932dynamical}. However, in recent years, with the rise of computer power and data science, this alternative framework of nonlinear dynamics analysis is extensively explored and a plethora of its applications are found in different fields~\cite{brunton2016koopman,arbabi2017ergodic,mezic2004comparison,mauroy2016global,mezic2005spectral}. In this new formulation, we focus on spectral properties of the Koopman operator, especially, the dominant eigenvalues and eigenfunctions. In high-dimensional systems, however, the Koopman operator is usually represented by a very large matrix which has numerous eigenvalues and a reliable way is still lacking on how to identify and interpret the dominant ones, not to mention an analytic approach to the problem, concerning its complexity.

Nevertheless, recent progress in the renormalization group (RG) theory may shed some light, which is capable of extracting global information from local expansion
~\cite{goldenfeld2018lectures}. Renormalization group investigates the change of physical laws across different scales and was first proposed for the perturbative calculation in quantum physics and statistical physics~\cite{zinn1996quantum}. Later, it was extended to the treatment of nonlinear dynamics for both flows and mappings~\cite{goldenfeld2018lectures,chen1994selection,paquette1994structural}. The starting point of the renormalization group is the removal of divergences (or resonance terms) from the perturbation series such that stable characteristics of system structure and dynamics are extracted which are insensitive to details, so it can be regarded as an altermative formulation of asymptotic analysis~\cite{goldenfeld2018lectures,chen1994selection,paquette1994structural}. Usually asymptotic analysis techniques such as multiple scale method (MS), boundary layer method (BL) and WKB approximation are sophisticated and quite daunting to use because of their complexity and limitations~\cite{vakakis1998analytic,kevorkian1982perturbation,bender2013advanced}. On the contrary, the renormalization group method does naive perturbation expansion and requires little prior knowledge and hence is very convenient to apply in practice. In Ref.~\cite{lan2013bridging}, the renormalization group analysis was further extended to the treatment of low-dimensional structures embedded in high-dimensional phase space. In the current paper, we will use renormalization group technique to carry out a perturbation analysis of synchronization process in the well-known Kuramoto model of coupled oscillators based on the Koopman operator.

Synchronization exists ubiquitously and is intensely studied, especially in coupled nonlinear systems, such as rhythmic applause of concert audience, synchronous vibration of pacemaker cells, fluid flow patterns, outbreak of infectious diseases, etc. Kuramoto model is a prototyped model for the study of synchronization and has become a paradigm~\cite{Zheng2012Phase,Rosenblum1996Phase,acebron2005kuramoto,chiba2018bifurcations,Wu2012Synchronizing,strogatz2000kuramoto}, though plenty of mysteries still exist concerning its dynamical behaviour. Thus, it constitutes a good test bed of the Koopman analysis for high-d nonlinear systems. Recently, Ott and Antonsen proposed a nice technique referred to as the QA method~\cite{ott2008low,ott2009long}, which rigorously defines a low-dimensional sub-manifold in the limit of large oscillator population that contains the synchronized state. Its success in diffrent contexts~\cite{martens2009exact,wu2016ott} indicates possible low-dimensional behaviour in the synchronization of networked systems though the dynamics could be weakly chaotic. Therefore, it seems possible to pin down possible synchronization transitions with single index defined in the Koopman analysis of the Kuramoto model even when the number of osciallators is finite.

Although the renormalization group approach is good for oscillator systems with weak coupling, due to the limitation in the expansion order or convergence, especially when the system size is large, it is hard to get the exact equation of motion for a real system with moderate or strong coupling. A lot of progress has been made in numerical analysis of the synchronization of Kuramoto model on complex networks, arousing a wave of research~\cite{chiba2018bifurcations,Wu2012Synchronizing,moreira2019global,zhang2019exponential}. In Ref.~\cite{chiba2018bifurcations}, global transition was found by checking the complex order parameter. But this method can not accurately get the critical coupling strength, not to mention the local synchronization point. In Ref.~\cite{Wu2012Synchronizing}, the local and global phase transition points can be found by plotting the frequency tree of the oscillation, but we still need to know the evolution of all oscillators. In this paper, we will use Koopman operator to analyze the Kuramoto model, based on the Hankel matrix and singular value decomposition (SVD) for dynamic mode decomposition (DMD). In view of the time delay embedding theory, we reconstruct local effective dynamical modes from a few oscillators. From the approximation of the Koopman operator, the frequency structure is identified and the important frequency component is obtained. At the same time, the phase transition point can be located by considering the correlation function of the dominant eigenfunctions at different coupling strengths. The reliability of the Koopman method will be verified in comparison with a direct numerical computation.

This paper is organized as follows: In Section~\ref{sec:koopman}, we will introduce Koopman operator, including its definition, eigenvalues and eigenfunctions of spectral decomposition, and commonly used algorithms for its computation. In Section~\ref{sec:nearestKuramoto}, we take the nearest-neighbor Kuramoto model as an example to carry out a perturbation analysis by a renormalization group method, which yields an approximate analytical solution. Then, the Koopman operator is applied to analyze the evolution data when the coupling is strong, in which eigenvalues and eigenfunctions are used to unfold the frequency structure of the system and locate the phase transition point. In section~\ref{sec:complexNetworksKuramoto}, the method is extended to the Kuramoto model on a complex network. The last section is the conclusion of the whole paper.
\section{Koopman operator\label{sec:koopman}}
Typical orbits are meandering around in a complex way in the phase space of a nonlinear system when the dynamics is chaotic, which poses considerable challenge to its description. Nevertheless, Koopman realized that the evolution of state-space observables can be described by a linear operator, which is later called the Koopman operator~\cite{koopman1931hamiltonian,koopman1932dynamical}.The overall dynamics could be decomposed into simpler modes characterized by its eigenvalues and eigenfunctions, which laid the foundation of a statistical approach to nonlinear evolution. Compared with the classical approach to individual state space trajectories, this method is more suitable for global analysis and optimal control. In this section, we will sketch the main ingredient of the theory and several numerical schemes for its application.
\subsection{Definition\label{sec:definition}}
Consider a continuous-time dynamical system in the $n$-dimensional phase space $\mathbb{M}$,
\begin{equation}
    \dot{x}=F(x)
    \,,
    \label{equ:continuousDynamical}
\end{equation}
where $x\in \mathbb{M}$ is the state vector and $F:\mathbb{M} \to \mathbb{M}$ gives the vector field which is usually nonlinear. $S^{t}(x_{0})$ denotes the solution to Eq.~\eqref{equ:continuousDynamical} with the initial value $x_{0}$ , which defines a differentiable flow in the phase space. For fixed $t$ , $S^{t}(x_{0})$ actually gives a diffeomorphism. The Koopman operator $\mathcal{K}$ is a linear operator, which acts on differentiable functions in the phase space of a dynamical system and gives the evolution of the function along orbits.

More explicitly, based on a differentiable function $g(x)$ in the phase space $\mathbb{M}$ , we may define its evolution
\begin{equation}
    g(t,x_{0})\equiv g(S^{t}(x_{0}))
    \,.
 \label{equ:GOnContinuousDynamical}
\end{equation}
All the observable functions constitute an infinite-dimensional linear vector space. A family of Koopman operators $\mathcal{K}^{t}$ ($t \in [0,\infty)$) may be defined in this linear space
\begin{equation}
    \mathcal{K}^{t} g(x_{0})=g(S^{t}(x_{0}))
    \label{equ:koopmanOnFlow}
    \,.
\end{equation}
For a fixed $t$ , $\mathcal{K}^{t}$ maps the observable $g(x)$ to $g(t, x)$ , which is actually more naturally defined in a discrete dynamical system. For a map $y=T(x)$ , the Koopman operator $\mathcal{K}$ acts on $g(x)$ as follows
\begin{equation}
\mathcal{K} g(x)=\tilde{g}(x)=g(T(x))
\label{equ:koopmanOnMap}
\,,
\end{equation}
which maps a function value to the next one on an orbit.
\subsection{\label{sec:egin}Eigenvalues and eigenfunctions}
From the above definition, it is easy to see that Koopman operator is closely related to the dynamics of the system. In fact, all the orbit information could be excavated from its action on a complete set of observables. Alternatively, a linear operator could also be characterized by its eigenfunctions and eigenvalues which may be more stable under perturbations and thus good for numerical computation. In Ref.~\cite{koopman1931hamiltonian}, Koopman himself related the spectral properties of the Koopman operator to conservativity, integrability and ergodicity. Later, Mezi{\' c} used eigenfunctions of Koopman operator to extract invariant sets and ergodic blocks in the state space~\cite{mauroy2016global,mezic2005spectral,budivsic2012geometry,mezic1999method,budivsic2009approximate,mezic2015applications}. Below, some spectral properties of the Koopman operator are discussed.

Let the eigenfunction $\varphi_{k}(x)$ of Koopman operator $\mathcal{K}$ correspond to the eigenvalue $\lambda_{k}$ , then
\begin{equation}
\begin{array}{cc}
  \mathcal{K} \varphi_{k}(x)=\varphi_{k}(T(x))=\lambda_{k}\varphi_{k}(x) \text{,} & k=1,2,\ldots
\end{array}
\label{equ:eginDefinition}
\,.
\end{equation}
If we have the eigenvalues $\lambda_{k_{1}}$, $\lambda_{k_{2}}$ with the eigenfunctions $\varphi_{k_{1}}(x)$, $\varphi_{k_{2}}(x)$, the product $\lambda_{k_{1}}\,, \lambda_{k_{2}}$ is also an eigenvalue with the eigenfunction $\psi(x)=\varphi_ {k_{1}}(x)\varphi_{k_{2}}(x)$, since
\begin{equation}
  \mathcal{K} \psi(x)=\psi(F(x))=\varphi_{k_{1}}(F(x))\varphi_{k_{2}}(F(x))=\lambda_{k_{1}}\lambda_{k_{2}} \varphi_{k_{1}}(x) \varphi_{k_{2}}(x)
  \,.
\end{equation}
Specifically, the $n$th ($n\in \mathbb{Z}$) power of $\lambda_{k}$ is also an eigenvalue, corresponding to the eigenfunction $\varphi_{k}^{n}(x)=(\varphi_{k}(x))^{n}$. Thus for the Koopman operator, producs of eigenfunctions are eigenfunctions, which may hinder the search for the principal eigenfunctions which look relatively simple but dominate the evolution in some sense.

The Koopman operator $\mathcal{K}$ can be uniquely decomposed into a regular and a singular part corresponding to the continuous and the discrete spectra~\cite{mezic2005spectral}. If the asymptotic dynamics is simple (fixed points, periodic orbits, invariant tori), the spectrum of the Koopman operator would have enough eigenvalues so that the eigenfunctions could be used as a basis to expand observables $g(x)$ as follows~\cite{mauroy2016global,arbabi2017ergodic}
\begin{equation}
  g(x)=\sum_{k}b_{k} \varphi_{k}(x)
  \,,
\end{equation}
where $b_k$ is the expansion coefficient. For a finite number of the basis functions, the expansion is approximate.\\
\indent On an ergodic component, the Koopman operator is unitary so its spectrum concentrates on a unit circle in the complex plane~\cite{mezic2005spectral}, while other eigenvalues mark the growth or decay of the corresponding modes and thus signal transient states. For $\lambda_{k}=1$ with the eigenfunction $\varphi_{k}(x)$, we have $\varphi_{k}(x_{p})=\mathcal{K} \varphi_{k}(x_{p-1})=\varphi_{k}(x_{p-1})=\ldots=\varphi_{k}(x_{0})$ ($p\in \mathbb{Z}$ represents the discrete time). Along an orbit, the eigenvalue does not change with time and all the points on the orbit assume the same value, which maps out an invariant subset of the dynamical system~\cite{budivsic2012geometry,mezic1999method}. For a long chaotic orbit, the subsect should be the ergodic component that contains the orbit. For $\lambda_{k}=e^{I \theta}$ ( $I=\sqrt{-1}$ is the imaginary unit.) where $\theta$ is real, if the corresponding eigenfunction $\varphi_{k}(x)$ robustly exists, then
\begin{equation}
\varphi_{k}(x_{p})=\mathcal{K} \varphi_{k}(x_{p-1})=e^{I \theta} \varphi_{k}(x_{p-1})=\ldots=e^{I p \theta}\varphi_{k}(x_{0})
\,,
\end{equation}
which depicts periodic motion of certain type~\cite{mezic2005spectral}.
\subsection{\label{sec:DMD}Dynamic Mode Decomposition}
With the development of computer and the great progress in numerical calculation, to obtain, process and analyze a large amount of dynamics data has become possible, and this data-driven approach enables new techniques and new paradigms~\cite{Kaiser2017Sparse}. The traditional point by point description of the system focuses on a single orbit, while the operator theory describes all orbits as a whole which thus naturally provides a global view of system dynamics. DMD is closely related to the spectral analysis of the Koopman operator~\cite{rowley2009spectral,Mezi2013Analysis}. In fact, even at the early times of its development~\cite{mezic2005spectral}, spectral decomposition is proposed for model reduction and mode decompositions based on data, either from experiments or simulation. DMD can be regarded as a numerical approximation of the Koopman spectral analysis. Below, different numerical schemes for the DMD will be introduced based on the Koopman operator.

Most often, a time series is available:
$$\bm{x_0},\bm{x_{\tau}},\bm{x_{2 \tau}},\ldots,\bm{x_{(n-1)\tau}},\bm{x_{n\tau}}\,,$$
which is used to extract key dynamical modes or information about the system under investigation. For this purpose, a dictionary of observables $g_{i}(\bm{x}) ,i=1,2,\ldots,m$ are selected to approximate the functional space, which leads to the following data matrix:
\begin{equation}
\begin{aligned}
  X= & \left( g_1(\bm{x}),g_2(\bm{x}),\ldots,g_m(\bm{x}) \right)\\
    =&  \left(
            \begin{array}{cccc}
                g_1(\bm{x_0}) & g_2(\bm{x_0}) & \ldots & g_m(\bm{x_0}) \\
                g_1(\bm{x_{\tau}}) & g_2(\bm{x_{\tau}}) & \ldots & g_m(\bm{x_{\tau}}) \\
                \vdots & \vdots & \ddots & \vdots \\
                g_1(\bm{x_{(n-1)\tau}}) & g_2(\bm{x_{(n-1)\tau}}) & \ldots & g_m(\bm{x_{(n-1)\tau}})
            \end{array}
        \right)
  \,,\\
  & \\
  Y= &\left( \tilde{g}_1(\bm{x}),\tilde{g}_2(\bm{x}),\ldots,\tilde{g}_m(\bm{x}) \right)\\
    =& \left(
            \begin{array}{cccc}
                g_1(\bm{x_{\tau}}) & g_2(\bm{x_{\tau}}) & \ldots & g_m (\bm{x_{\tau}}) \\
                g_1(\bm{x_{2\tau}}) & g_2(\bm{x_{2\tau}}) & \ldots & g_m(\bm{x_{2\tau}}) \\
                \vdots & \vdots & \ddots & \vdots \\
                g_1(\bm{x_{n\tau}}) & g_2(\bm{x_{n\tau}}) & \ldots & g_m(\bm{x_{n\tau}})
            \end{array}
        \right)
    \,,
\end{aligned}
\label{equ:dataMatrix}
\end{equation}
where each column of the data matrix $X$, $Y$ is an $n$-point representation of the observables and the points are picked up uniformly in time from a long orbit. The number $n$ of representative points should be large enough to give good spatial resolution. The projection $\mathcal{K}^{\tau}$ of the Koopman operator in this finite-dimensional function space, therefore, could be obtained from the equation below:
\begin{equation}
    \mathcal{K}^{\tau}X=Y\,,\mbox{ or } \mathcal{K}^{\tau}=YX^T(XX^T)^{-1}
    \,,
    \label{equ:koopman}
\end{equation}
by a matrix inversion (or pseudo-inversion). The selection of basis functions is very important, which depends on the features to be explored. We need enough basis functions to describe these features and reach reliable conclusions~\cite{brunton2016extracting,rowley2009spectral,schmid2010dynamic}. For a good basis, $m$ could be small and the computation load is thus much reduced, especially for high-dimensional systems. But an optimal selection of the truncation and the form of basis functions is a difficult problem.

Commonly used basis functions include Gaussian basis, Fourier basis, polynomial basis and so on, which may be good on certain occasions but may not be so on others. Fortunately, it is possible to numerically extract a natural basis based on the Hankel matrix, which directly uses the delay embedding of the measurement values, {\em i.e. $g(x)=x$} . Delay embedding is a geometric reconstruction scheme for attractors in nonlinear systems and extremely useful in the modern theory of time series analysis~\cite{sauer1991embedology,takens1981detecting}. In Ref.~\cite{arbabi2017ergodic}, it is proved that the eigenfunctions and eigenvalues of Koopman operator can be obtained by applying DMD to the Hankel data matrix with limited observations. By combining delay embedding with DMD, we are able to extract interesting dynamics information about the inaccessible state space from the partial data. More explicitly, we write the Hankel matrix as follows:
\begin{equation}
 \bm{H}=\left(
  \begin{array}{cccc}
    \bm{x_{0}} & \bm{x_{\tau}} & \ldots & \bm{x_{m\tau}} \\
    \bm{x_{\tau}} & \bm{x_{2\tau}} & \ldots & \bm{x_{(m+1)\tau}} \\
    \vdots & \vdots & \ddots & \vdots \\
    \bm{x_{(n-1)\tau}} & \bm{x_{n\tau}} & \ldots & \bm{x_{(n+m-1)\tau}}
  \end{array}
  \right)
  \,,
\end{equation}
where each column could be viewed as an observable based on the $n$ points along a typical trajectory in the phase space. As is well known, a non-trivial smooth function defined in the phase space will soon become rather rugged upon the action of the Koopman operator if the system is chaotic. Therefore, the high wave-number undulations could not be captured in this representation and the best that we can hope for is a good description of features (or averages of features) which could be depicted with the $n$ points. As the column vectors of the Hankel matrix are usually not even close to orthogonal, it is a good practice to first apply the SVD to the Hankel matrix to extract important directions forming an orthogonal basis~\cite{brunton2017chaos} . As a result, the large-scale structures are contained in the first, say, $r$ columns. We follow this procedure in the analysis below by writing the Hankel matrix as
\begin{equation}
\begin{aligned}
  \bm{H}& = \bm{U \Sigma V^{T}}     \\
        & = \left(
            \begin{array}{cccc}
            | & | & & | \\
            \bm{u_{1}} & \bm{u_{2}} & \ldots & \bm{u_{m}} \\
            | & | & & |
            \end{array}
            \right)
            \left(
            \begin{array}{cccc}
            \sigma_{1}  & 0             & \ldots & 0 \\
            0           & \sigma_{2}    & \ldots & 0 \\
            \vdots      & \vdots        & \ddots & \vdots \\
            0           & 0             & \ldots & 0
            \end{array}
            \right)
            \left(
            \begin{array}{cccc}
            -& \bm{v_{1}^{T}} & -\\
            -& \bm{v_{2}^{T}} & -\\
             & \vdots &  \\
            -& \bm{v_{n}^{T}} & -\\
            \end{array}
            \right)
\end{aligned}
\,,
\label{equ:SVD}
\end{equation}
where as a common practice the positive singular values are arranged in a decreasing order, reflecting the relative importance of the corresponding columns of $U$ and $V$. The columns of $\bm{U}$ could be viewed as a new orthogonal basis while each column of $\bm{V}^T$ is the coordinates of the observable in this new basis divided by the corresponding singular value. In another word, the original Hankel matrix becomes $\Sigma\bm{V}^T$ in the new frame. The selection of the truncation order $r$ depends on the relative magnitude of the singular values as well as the features under investigation. For safe, here we keep all the modes with $\sigma_i \geq 10^{-7} $ and thus
\begin{equation}
\begin{aligned}
  \bm{\acute{H}}    & =\bm{\acute{U} \acute{\Sigma} \acute{V}^{T}} \\
                    & =\left(
                        \begin{array}{cccc}
                        | & | & & | \\
                        \bm{u_{1}} & \bm{u_{2}} & \ldots & \bm{u_{r}} \\
                        | & | & & |
                        \end{array}
                        \right)
                        \left(
                         \begin{array}{cccc}
                        \sigma_{1}  & 0             & \ldots & 0 \\
                        0           & \sigma_{2}    & \ldots & 0 \\
                        \vdots      & \vdots        & \ddots & \vdots \\
                        0           & 0             & \ldots & \sigma_{r}
                        \end{array}
                        \right)
                        \left(
                        \begin{array}{cccc}
                        -& \bm{v_{1}^{T}} & -\\
                        -& \bm{v_{2}^{T}} & -\\
                        & \vdots &  \\
                        -& \bm{v_{r}^{T}} & -\\
                        \end{array}
                        \right)
\end{aligned}
\,.
\label{equ:SVDDimensionReduction}
\end{equation}
A Koopman analysis based on $\bm{\acute{\Sigma}\acute{V}}$ could be performed similar to what has been done in Eq.~\eqref{equ:koopman} .
The eigenfunctions could be represented in the original function space by a matrix multiplication. As non-essential parts are ignored, the computational complexity is greatly reduced. As a result, we achieve both efficiency and stability in the computation. The SVD is essentially a linear transformation, which is widely used in numerical analysis for dimension reduction or noise filtering~\cite{brunton2017chaos,rowley2005model}. In the current context, the Koopman analysis will further reduce the dimension by embedding the temporal evolution in the analysis while maintaining the robustness by keeping only the dominant eigenfunctions.
\section{\label{sec:nearestKuramoto}Application to a Kuramoto model with nearest-neighbor coupling}
Synchronization is closely related to our life, so it has always been a hot research topic~\cite{pikovsky2003synchronization} . In 1975 , Kuramoto proposed his model~\cite{winfree1967biological} based on Winfree's research~\cite{kuramoto1975international} , which is classic for the description of synchronous behavior of a large number of coupled oscillators. It was originally used in biological rhythm, chemical oscillations, and later extended to other applications~\cite{acebron2005kuramoto,chiba2018bifurcations,Wu2012Synchronizing,strogatz2000kuramoto} , which describes limit-cycle oscillators with distinct natural frequencies and sinusoidal couplings, where amplitude information is ignored and only phase information is considered. The model is simple enough for various analytical derivation, rich enough to display a variety of synchronization modes, and flexible enough to adapt to a variety of physical scenarios.

Without lack of generality, we consider the classical and simplest Kuramoto model with a general coupling structure:
\begin{equation}
\label{equ:KMGeneral}
\begin{array}{cc}
  \dot{\theta}_{i}=\omega_{i}+K \sum\limits_{j=1}\limits^{N} a_{ij} \sin(\theta_{j}-\theta_{i}) \text{,} & i=1,\cdots,N
\end{array}
\,,
\end{equation}
where $N$ is the total number of oscillators. $\theta_{i}$ is the phase of the $i$-th oscillator and $\omega_{i}$ is the natural frequency of the $i$-th oscillator drawn from a common distribution $\rho(\omega)$. The parameter $K$ is the coupling strength (also known as coupling coefficient) between oscillators. $a_{ij}=0 \text{ or }1$ indicates the adjacency between oscillators. In the original Kuramoto model, $a_{ij}=1$ for all pairs, featuring a complete interaction graph. Later on, various coupling strategies are proposed and studied on different networks.
\subsection{Kuramoto model on a circle}
In this section, we consider the nearest-neighbor Kuramoto model~\cite{zheng1998phase,kogan2009renormalization,el2009transition}, which includes only nearest-neighbor couplings on a circle:
\begin{equation}
\label{equ:KMNearest}
\begin{array}{cc}
  \dot{\theta}_{i}=\omega_{i}+K(\sin(\theta_{i+1}-\theta_{i})+\sin(\theta_{i-1}-\theta_{i}))\text{,} & i=1,\cdots,N
\end{array}
\,.
\end{equation}
Here, the circle topology indicates a periodic boundary condition: $\theta_{0}=\theta_{N}$, $\theta_{N+k}=\theta_{k}$.

It can be seen from Eq.~\eqref{equ:KMNearest} that an oscillator only interacts with its two neighbours in an attractive manner. When the coupling is weak or the natural frequency difference is large, the oscillators tend to rotate independently and the overall dynamics is incoherent. With increasing coupling strength, the average frequency of neighboring oscillators tends to condense. When a threshold is reached, a small group of oscillators fall into synchrony and more or less acts like one "big" oscillator. For large enough coupling strength, all oscillators get synchronized and rotate together with a frequency which is the algebraic average of the natural frequencies of these $N$ oscillators. To characterize the long-time rotation behaviour, an average frequency is defined for each oscillator $i$ :
\begin{equation}
\hat{\omega}_{i}=\langle\dot{\theta}_{i}\rangle=\lim\limits_{T\rightarrow\infty}\frac{1}{T}\int_{0}^{T}\dot{\theta}_{i}(t)dt
\,,
\end{equation}
which is a relatively simple index and amenable to numerical computation but may be used to select important modes in the Koopman analysis as will be shown below.

To facilitate the analytic computation, the change of the variable $z_{i}=e^{I\theta_{i}}$ ( $I=\sqrt{-1}$ is the imaginary unit) in Eq.~\eqref{equ:KMNearest} entails a polynomial vector field:
\begin{equation}
\begin{array}{cc}
  \dot{z}_{i}=I\omega_{i}z_{i}+\frac{K}{2}(z_{i+1}-{z_{i}}^{2}\bar{z}_{i+1}+z_{i-1}-{z_{i}}^{2}\bar{z}_{i-1})\text{,} & i=1,\cdots,N
\end{array}
\label{equ:KMExponential}
\,,
\end{equation}
where the bar over a symbol denotes complex conjugation. The boundary condition then becomes $z_0=z_N$, $z_{N+k}=z_{k}$.

In the following, without loss of generality, Koopman analysis will be carried out for Eq.~\eqref{equ:KMExponential} with $N=6$ oscillators, and its natural frequencies are randomly selected as: $-1$ , $-0.7$ , $-0.4$ , $0$ , $0.2$ , $0.6$ . In the limit of weak coupling, a perturbation technique based on the renormalization group analysis could be designed to compute the eigenfrequency and the eigenfunction. Even the lowest order approximation matches well with the numerical computation. If the coupling becomes strong, the analytic approach becomes overly cumbersome and is relegated to pure numerical calculation.
\subsection{Renormalization group analysis} \label{sec:RG}
After Goldenfeld {\em et al} extended the RG analysis to the treatment of differential dynamical systems, great progress has been made in both applications and theoretical understanding. In fact, many singular perturbation methods can be understood in terms of a renormalization process. The amplitude or phase equations could be regarded as RG equations and thus derivable in the framework of the RG theory~\cite{chen1994renormalization}. A geometric interpretation is proposed based on the classical theory of envelopes in differential geometry~\cite{kunihiro1995geometrical,kunihiro1997geometrical,ei2000renormalization,hatta2002renormalization,kunihiro1998renormalization1,kunihiro1998renormalization2,kunihiro1998dynamical,kunihiro2006application}, which leads to applications in the computation of  central manifold~\cite{Chiba2008Approximation}, heteroclinic orbit in nonlinear systems~\cite{lan2013bridging}, and so on. Here, we apply the RG technique to a perturbation analysis of the nearest-neighbor Kuramoto model~\eqref{equ:KMExponential}.

The RG analysis starts from a naive series expansion of the solution to the differential equation. In fact, an oscillator rotates with an average frequency after transient, rather than its natural frequency. Hence, we introduce a new parameter $\tilde{\omega}_{i}$ indicating the average frequency of the $i$-th oscillator and get the transformed formula:
\begin{equation}
  \dot{z}_{i}=I\tilde{\omega}_{i}z_{i}+\frac{\varepsilon K}{2}(z_{i+1}-{z_{i}}^{2}\bar{z}_{i+1}+z_{i-1}-{z_{i}}^{2}\bar{z}_{i-1})-I\varepsilon^{2} (\triangle \omega_{i}) z_{i}
  \,,
  \label{equ:transformedKmWithAverageFrequency}
\end{equation}
where $\triangle \omega_{i}=\tilde{\omega}_{i}-\omega_{i} $ , being an unknown frequency correction term of the $i$-th oscillator, which may be used to eliminate possible resonance terms.

Next, we make an expansion in $\varepsilon$:
\begin{equation}
\begin{split}
z_{i}&=z_{i0}+\varepsilon z_{i1}+\varepsilon^{2} z_{i2}+\cdots
\,,\\
\triangle \omega_{i} &=\triangle \omega_{i0}+\varepsilon \triangle \omega_{i1}+\varepsilon^{2} \triangle \omega_{i2}+\cdots
\,,
\end{split}
\end{equation}
where the first subscript $i$ in $z_{ik}$ and $\triangle \omega_{ik}$  represents the index of the oscillator, and the second subscript $k$ indicates the expansion order.
Put the perturbation expansion into Eq.~\eqref{equ:transformedKmWithAverageFrequency} and compare different powers of $\varepsilon$ , we get a set of differential equations:
\begin{widetext}
\begin{equation}
\begin{aligned}
\varepsilon^{(0)} \text{:} \dot{z}_{i0}=& I \tilde{\omega}_{i} z_{i0} \\
\varepsilon^{(1)} \text{:} \dot{z}_{i1}=& I  \tilde{\omega}_{i}z_{i1}+\frac{K}{2}(z_{(i+1)0}-{z_{i0}}^{2}\bar{z}_{(i+1)0}+z_{(i-1)0}-{z_{i0}}^{2}\bar{z}_{(i-1)0}) \\
\varepsilon^{(2)} \text{:} \dot{z}_{i2}=& I \tilde{\omega}_{i} z_{i2}+\frac{K}{2}(z_{(i+1)1}-{z_{i0}}^{2}\bar{z}_{(i+1)1}-2 z_{i0} z_{i1} \bar{z}_{(i+1)0} +z_{(i-1)1}-{z_{i0}}^{2}\bar{z}_{(i-1)1}-2 z_{i0} z_{i1} \bar{z}_{(i-1)0}) -I \triangle \omega_{i0} z_{i0}   \\
   \cdots&
\end{aligned}
\end{equation}
\end{widetext}
The first equation is a linear differential equation, which can easily be solved.
\begin{equation}
z_{i0}=A_{i}e^{I \tilde{\omega}_{i} (t-t_{0})}
\,,
\end{equation}
where $A_{i}$ is the initial value, being a function of $t_{0}$ , to be used as the renormalization variable later.

The higher-order equations can be solved based on low-order results. For example, at the first order, to eliminate the resonance term, we set $\triangle \omega_ {i0}=\frac{K^{2}}{\tilde{\omega}_{i-1}-\tilde{\omega}_{i}}+\frac{K^{2}}{\tilde{\omega}_{i+1}-\tilde{\omega}_{i}}$ leading to:
\begin{widetext}
\begin{equation}
z_{i1}= \frac{K}{2} \frac{I}{\tilde{\omega}_{i}-\tilde{\omega}_{i-1}} ( z_{(i-1)0}+ \bar{z}_{(i-1)0} z_{i0}^{2} )+
\frac{K}{2} \frac{I}{\tilde{\omega}_{i}-\tilde{\omega}_{i+1}}( z_{(i+1)0}+ z_{i0}^{2} \bar{z}_{(i+1)0} )
\,.
\end{equation}
\end{widetext}
Because of the limited space, higher order results will not be listed. The approximate solution is:
\begin{widetext}
\begin{equation}
  \acute{z}_{i}=  z_{i0}+ \frac{\varepsilon K}{2} \left[\frac{I}{\tilde{\omega}_{i}-\tilde{\omega}_{i-1}}(z_{(i-1)0}+ \bar{z}_{(i-1)0} z_{i0}^{2} )\right.
   \left. + \frac{I}{\tilde{\omega}_{i}-\tilde{\omega}_{i+1}}(  z_{(i+1)0}+ z_{i0}^{2} \bar{z}_{(i+1)0} ) \right]+\cdots
\label{equ:solveResult}
\,,
\end{equation}
\end{widetext}
where $\acute{z}_{i}=\acute{z}_{i}[t;t_{0},\bm{A(t_{0})}]$ is a function of the initial time $t_{0}$ and the initial vector $\bm{A(t_{0})}$ . The higher the order of ${\varepsilon}$, the more accurate is the approximate analytic solution.

We can get the polynomial expansion of $A_i$ on $z_i$ from Eq.~\eqref{equ:solveResult} after setting $t=t_0$ :
\begin{widetext}
\begin{equation}
 A_{i}=\frac{c_i}{2} z_{i-1}
        + (1+ \frac{c_i^2}{2}+\frac{ (-c_{i+1})^2 }{2} ) z_i
        + \frac{ -c_{i+1} }{2} z_{i+1}
        +\frac{ c_i }{2} \bar{z}_{i-1} z_i^2
        +\frac{ -c_{i+1} }{2} z_i^2 \bar{z}_{i+1}
        -\frac{c_i^2}{8} z_{i-1}^2 \bar{z_i}
        -\frac{(-c_{i+1})^2}{8} \bar{z_i} z_{i+1}^2
        +\dots
\,,
\label{equ:expandsionOfAOnZ}
\end{equation}
\end{widetext}
where $c_i=\frac{ K \varepsilon}{ I(\tilde{\omega}_i-\tilde{\omega}_{i-1}) }$ .

The RG principle tells that the physical quantity should be independent on the starting point of the renormalization~\cite{chen1994renormalization}. Adopting the idea from Ref.~\cite{lan2013bridging}, we focus on the dynamics induced by the coordinate $z_i$ on some submanifold, along the eigen-direction in the complex domain near the origin.
The initial vector is $\bm{A}=(0,\dots,A_{i},\dots,0)$, emphasizing the analytic structure started from the $A_{i}$ component. According to the renormalization equation
\begin{equation}
\frac{d \acute{z}_{i}[t;t_{0},A_{i}(t_{0})] }{ d t_{0}} \Big| _{t=t_{0}}=0
  \,,
  \label{equ:RFunction}
\end{equation}
the equation for $d A_{i}(t_{0})/d t_{0}$ could be obtained. To remove non-autonomous terms, we take a limit $t=t_{0}$ . The resulting evolution equation for $A_{i}(t_{0})$ is
\begin{equation}
  \dot{A}_{i}=i \tilde{\omega}_{i} A_{i}
  \label{equ:RGFunction}
  \,,
\end{equation}
which looks simple and its solution is $A_{i}(t_0)=A_{i,0}e^{i\tilde{\omega}_{i}t_0}$ , where $A_{i,0}$ is the initial value. Based on Eq.~\eqref{equ:koopmanOnFlow} in Sec.~\ref{sec:definition} and Eq.~\eqref{equ:eginDefinition} in Sec.~\ref{sec:egin} , $\mathcal{K}^{\tau} A_{i,0}=A_{i}(\tau)=A_{i,0}e^{i\tilde{\omega}_{i} \tau}$ , which indicates that $e^{i \tilde{\omega}_{i} \tau}$ and $A_{i,0}$ (or $A_i$) is an eigenvalue and eigenfunction of the Koopman operator $\mathcal{K}^{\tau}$ respectively. Here, we keep the simple oscillation dynamics by eliminating the resonance terms order by order, and it turns out that the renormalized frequency $\tilde{\omega}_i$ matches very well with the average frequency $\hat{\omega}_i$ computed from the time series.
\subsection{Koopman analysis}  \label{sec:koopmanForLine}
\begin{table}
\caption{Comparison of coefficients of the principal eigenfunction obtained by the RG and the Koopman computation for the Kuramoto model on a circle of $N=6$ oscillators at $K=0.08$ . The last line is the ratio of the two.}
\begin{center}
\begin{ruledtabular}
\begin{tabular}{ccccc}
         & $z_{1}$  & $z_{2}$ & $z_{1}^{2} \bar{z_{2}}$ & $\bar{z_{1}}z_{2}^{2}$ \\
\hline
RG     & 1.00000     & 0.14588   & 0.14588   &0.01064 \\
Koopman    & 43.98444	&5.84375	&5.79867	&0.42083\\
RG /Koopman    & 0.02274    &0.02496 &0.02516	&0.02529\\
\end{tabular}
\end{ruledtabular}
\end{center}
\label{tab:N6ExpandCoefficientCompare}
\end{table}
\begin{figure*}
  \centering
   \subfigure[\quad $K=0.61$]{ \label{fig:spectrumbase1Sub1}
    \begin{minipage}[b]{0.45\textwidth}
    \centering
    \includegraphics[width=1\textwidth]{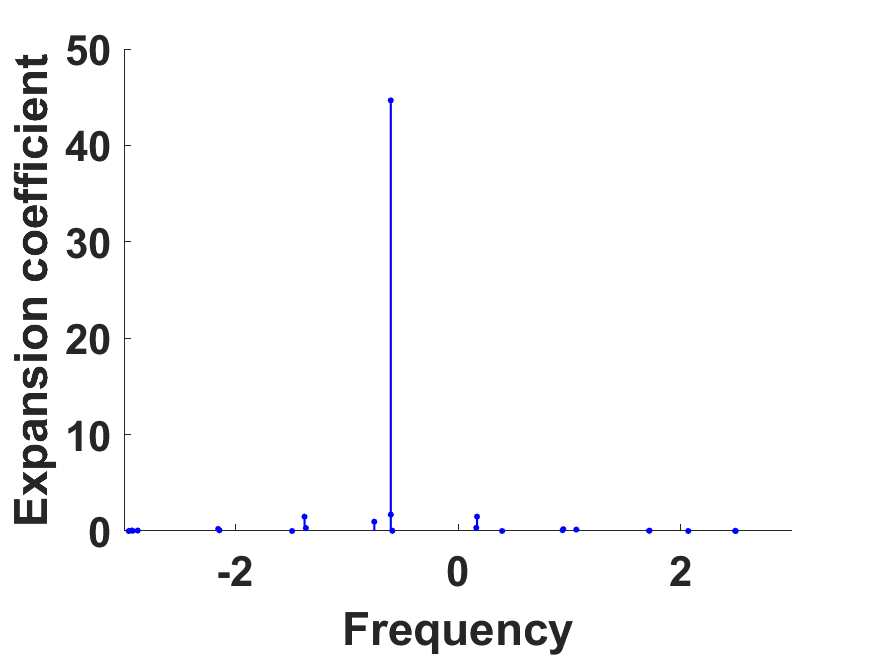}
    \end{minipage}
  }
   \subfigure[\quad $K=1.21$]{ \label{fig:spectrumbase1Sub2}
    \begin{minipage}[b]{0.45\textwidth}
    \centering
    \includegraphics[width=1\textwidth]{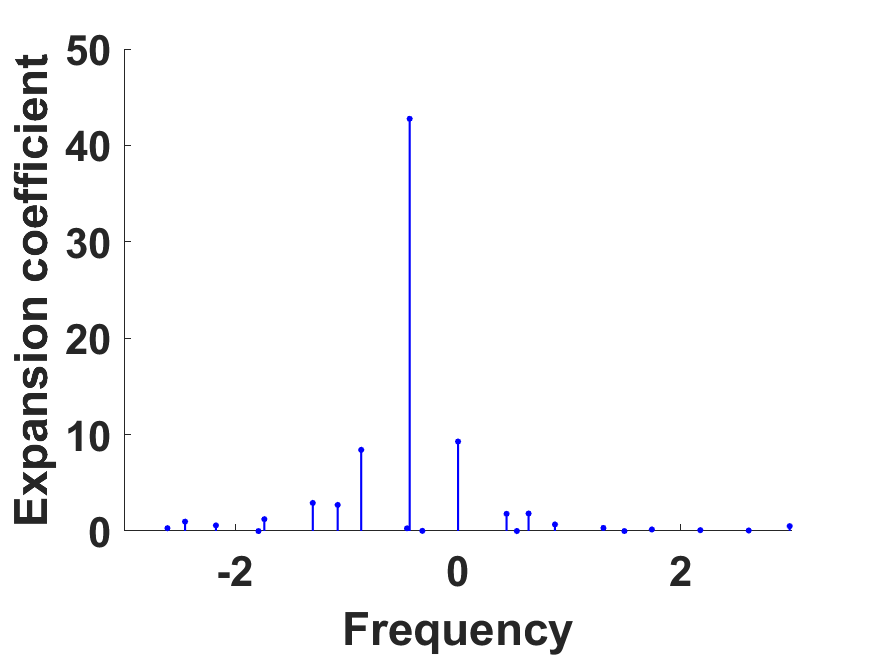}
    \end{minipage}
  }

   \subfigure[\quad $K=1.43$]{ \label{fig:spectrumbase1Sub3}
    \begin{minipage}[b]{0.45\textwidth}
    \centering
    \includegraphics[width=1\textwidth]{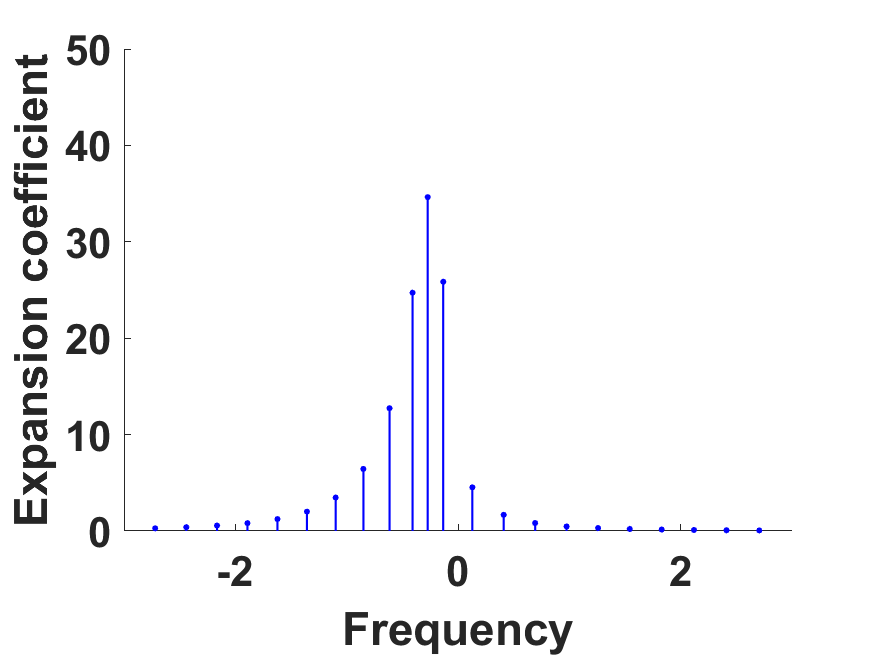}
    \end{minipage}
  }
   \subfigure[\quad $K=1.45$]{ \label{fig:spectrumbase1Sub4}
    \begin{minipage}[b]{0.45\textwidth}
    \centering
    \includegraphics[width=1\textwidth]{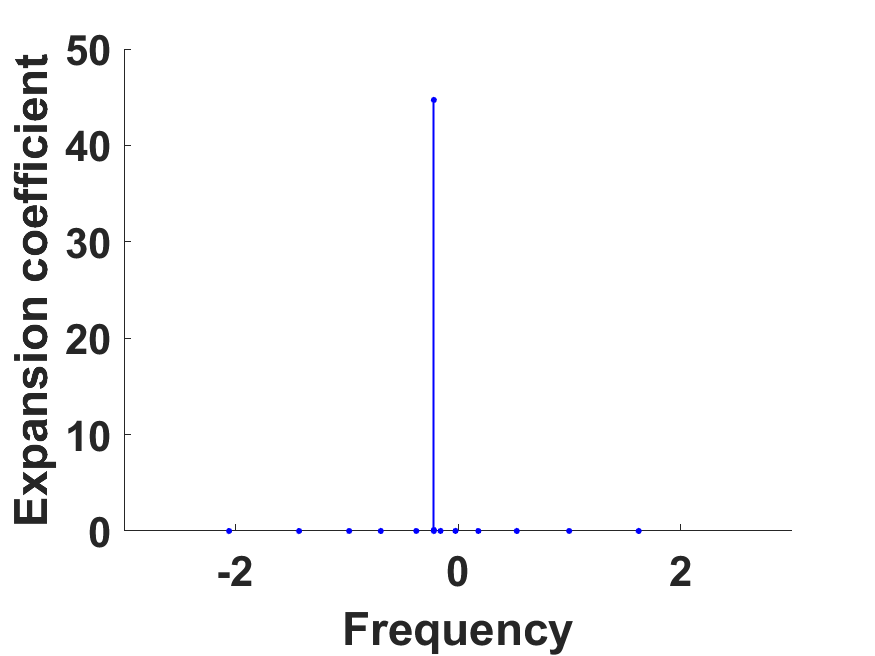}
    \end{minipage}
  }
    \caption{Expansion coefficients of the observable $(z_{1}^{0},z_{1}^{\tau},\dots,z_{1}^{(n-1)\tau})^{T}$ on the eigenfunctions of the Koopman operator for the Kuramoto model on a circle of $N=6$ oscillators with different coupling strengths when $\bm{Z}^{t}=(z_{1}^{t})$ .  }
  \label{fig:spectrumbase1}
\end{figure*}
The above analytic approach is possible only for lowest orders and converging well for weak couplings. In order to study more general cases, numerical computation has to chip in. In this section, we will analyze the frequency structure of the coupled system, then extract dominant frequencies and locate phase transition points according to the eigenvalues of the Koopman operator.

We use the observable(s) $\bm{Z}^{t}$ to construct the Hankel matrix:
\begin{equation}
 \bm{H}=\left(
  \begin{array}{cccc}
    \bm{Z^{0}} & \bm{Z^{\tau}} & \ldots & \bm{Z^{m\tau}} \\
    \bm{Z^{\tau}} & \bm{Z^{2\tau}} & \ldots & \bm{Z^{(m+1)\tau}} \\
    \vdots & \vdots & \ddots & \vdots \\
    \bm{Z^{(n-1)\tau}} & \bm{Z^{n\tau}} & \ldots & \bm{Z^{(n+m-1)\tau}}
  \end{array}
  \right)
  \,,
\end{equation}
where $\bm{Z}^{t}$ can be a row or a column vector composed of $z_{i}^t=z_{i}(t)$ , and the superscript $t$ is the time indicator. According to the algorithm in Sec.~\ref{sec:DMD}, with the evolutionary step $\tau=0.01$, $n=2000$ and $m=5000$ , a finite dimensional approximation of the Koopman operator can then be obtained, available for the computation of eigenvalues and eigenfunctions.

Here, let's take $\bm{Z}^{t}=(z_{1}^{t})$ , only the observations from oscillator 1. In Fig.~\ref{fig:spectrumbase1} , we draw the expansion coefficient of the observable
\begin{equation}
    (\bm{Z}^{0},\bm{Z}^{\tau},\dots,\bm{Z}^{(n-1)\tau})^{T}=(z_{1}^{0},z_{1}^{\tau},\dots,z_{1}^{(n-1)\tau})^{T}
\end{equation}
with respect to eigenfunctions ordered by frequencies with different coupling strengths. For ease of plotting, we limit the abscissa to the interval $[-3,3]$. The figure is similar to the spectrum diagram and is discrete and asymmetric. Obviously, the most important eigenvalue is related to the average frequency of oscillator 1 as depicted in Sec.~\ref{sec:RG} . According to Sec.~\ref{sec:egin}, the linear combination of the average frequencies is also related to eigenvalues, and the corresponding eigenfunctions can also be recovered. From Fig.~\ref{fig:spectrumbase1Sub1} to Fig.~\ref{fig:spectrumbase1Sub3}, it can be observed that the components become more and more complex with the increase of the coupling strength, and the frequency proportion of nonlinear components increases steadily. Of course, some local synchronizations in the process are not clearly observed from the figure. When reaching a critical coupling strength, all oscillators are synchronous and experiencing a global phase transition, where the main frequency is the average of individual generic frequencies as shown in Fig.~\ref{fig:spectrumbase1Sub4}. It seems that the system can be treated as "one" cluster rotating at a common frequency to some extent after synchronization. To sum up, eigenfunctions describe modes with different frequencies, and the expansion coefficients on the eigenfunctions of the observable give a kind of "spectrum". From the change of spectrum with the coupling strength, we find the global synchronization point $K=1.45$ with an error of about 0.02.

In Tab.~\ref{tab:N6ExpandCoefficientCompare} , we compare of the principal eigenfunction obtained by the Koopman and the RG computation (2nd-order approximation) for $K=0.08$. The ratio of the corresponding coefficients is about 0.025, which indicates that the two results are proportional to each other and the two approaches well match in case of small couplings.
\subsection{Tracking of bifurcations}
\begin{figure}
    \centering
    \includegraphics[scale=0.6]{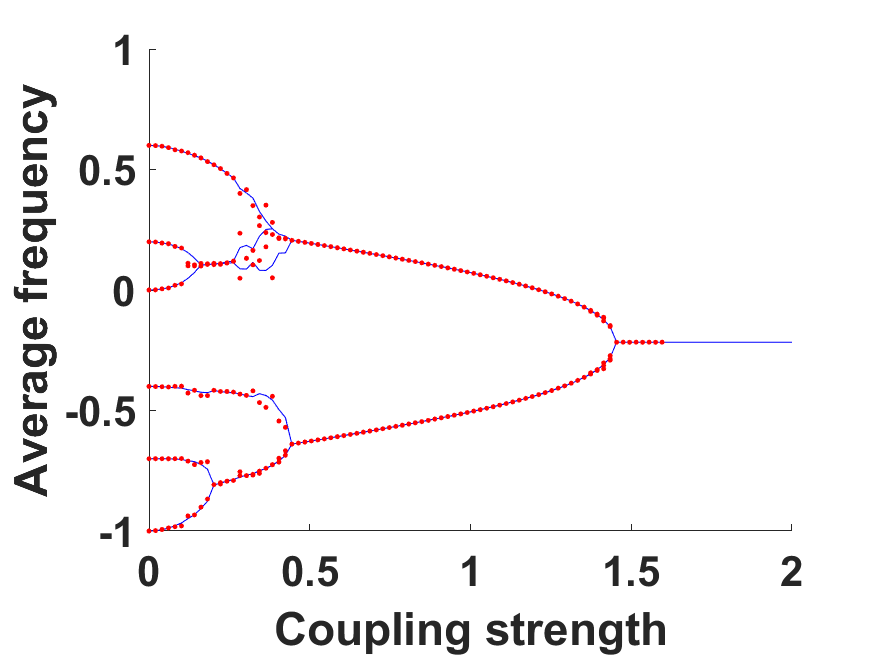}
    \caption{ The average frequency {\em v.s.} coupling strength $K$  for the Kuramoto model on a circle of $N=6$ oscillators when $\bm{Z}^{t}=(z_{i}^{t}),i=1,2,\dots,6$. The blue curve is the numerical solution, and the red hollow circle is the Koopman solution.}
    \label{fig:N6NearestBifurcationBase1}
\end{figure}
\begin{figure}
  \centering
    \includegraphics[scale=0.6]{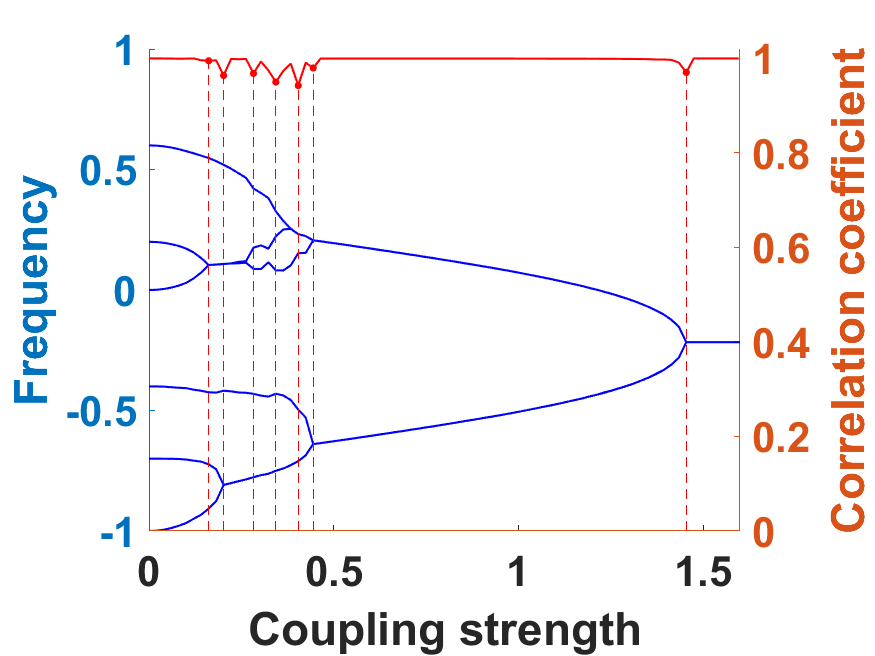}
  \caption{Correlation coefficient (red dashed line) and the average frequency (blue solid lines) {\em v.s.} the coupling strength $K$ for the Kuramoto model on a circle of $N=6$ oscillators when $\bm{Z}^{t}=(z_{i}^{t}),i=1,2,\dots,N$.}
  \label{fig:correlationCoefficientN6}
\end{figure}
\begin{table}
\centering
\caption{The critical coupling strength for the Kuramoto model on a circle of $N=6$ oscillators obtained by the direct numerical and the Koopman computation.}
\begin{ruledtabular}
\begin{tabular}{cccccccc}
numerical   &0.16&0.20&0.26&0.32&0.38&0.44&1.45\\
Koopman &0.16&0.20&0.28&0.34&0.40&0.44&1.45\\
\end{tabular}
\end{ruledtabular}
\label{tab:N6PhaseTransitionCompare}
\end{table}
As the coupling strength increases, the actively involved eigen-modes increase first, at some point start to decrease and collapse to a principal one beyond the critical point. How to find the important ones and how to determine the bifurcation points along the way are key questions that we need to check within the framework of the Koopman operator. Through the analysis above, we find that no matter what the coupling strength is, the largest expansion coefficient is always attached to the average frequency which is thus regarded as the principal mode of the oscillator.

If we take the observable $\bm{Z} ^{t}=(z_{i}^{t}) , i=1,2,\dots,N$ , the eigenfunction expansion of the motion of each oscillator is different and changes with coupling strength, while the dominant eigenfunctions always correspond to the average frequencies of individual oscillators. In virture of this observation, we may extract six sets of spectra for different coupling strengths but with $N=6$ as depicted in Fig.~\ref{fig:N6NearestBifurcationBase1}. Blue solid lines display the numerical result, and red hollow circles for the Koopman solution. However, the Koopman solution at the local phase transition point may not be very accurate, because upon synchronization, the motion of the involved oscillators is very sensitive to the coupling strength or perturbations from other degrees of freedom. Further investigation is needed in the future for a more accurate computation near the bifurcation point. On the whole, however, the Koopman results match the numerical ones well. When the observable is a row vector, $\bm{Z}^{t}=(z_{i}^{t},z_{i+1}^{t})$ or a column vector $\bm{Z}^{t}=(z_{i}^{t},z_{i+1}^{t})^{T}$ , the koopman analysis gives a result similar to what is in Fig.~\ref{fig:N6NearestBifurcationBase1} so we will not show it again.

From the above discussion we know that there is a correspondence between the eigenvalues and the average frequencies of the oscillators, while the eigenfunctions map out relative importance of different modes in the phase space.Thus, it is possible to find out the phase transition point by checking the change of the eigenfunction with the coupling strength. With $\bm{Z} ^{t}=(z_{n}^{t})$ , we may define the correlation coefficient of the eigenfunction $\varphi_{n}$ of oscillator n:
\begin{equation}
  \rho_{n}(K_{j})=\frac{Cov(\varphi_{n} (K_{j-1}),\varphi_{n} (K_{j}))}{\sqrt{Var(\varphi_{n} (K_{j-1}))Var(\varphi_{n} (K_{j}))}}
  \,,
\end{equation}
where in this section, we take $K_{j}-K_{j-1}=0.01$. The eigenfunction $\varphi_n(K_j)$ receives the largest expansion coefficient of the observable at certain coupling strength $K_j$ , which can be used to predict some of the synchronization points. For better characterization, we take the summation: $\rho=\frac{1}{N} \sum_{n=1}^{N} \rho_{n}(K_{j})$ , and draw the "$\rho$-$K$" graph with the red dashed line in Fig.~\ref{fig:correlationCoefficientN6} . In most cases, the correlation coefficient $\rho$ is close to 1, because the mode changes smoothly with the coupling strength if there is phase transition. When a local or a global synchronization is about to occur, there is a rapid change of the principal eigenvalue and eigenfunction, resulting in a perceivable change of $\rho$. In the graph, henceforth, a dip in a smooth background is featured for $\rho \textless 1$ . By comparing with the numerical result in Tab.~\ref{tab:N6PhaseTransitionCompare}, we find that the error in the prediction of synchronization point is within 0.02.

It can be seen that the Koopman method only needs evolution data, but is able to extract important frequencies through the eigenvalues and eigenfunctions. Based on the correlation variation with the coupling strength, multiple phase transition points can be detected. In a word, in the nearest-neighbor Kuramoto model, the numerical solution, the analytical solution of the RG method with small coupling and the results of the Koopman analysis match each other well, which proves the reliability of the Koopman method.
\section{\label{sec:complexNetworksKuramoto}Application to a Kuramoto model on a complex network}
In the above section, Koopman operator is used to analyze the nearest-neighbor Kuramoto model of $N=6$ oscillators. But in reality, the coupled system is usually defined on a complex network, which is a hot research topic in recent years~\cite{chiba2018bifurcations}. In this section, we will extend our application of the Koopman analysis to  the Kuramoto model on a complex network. First, we discuss the network topology.
\subsection{Kuramoto model on a complex network }
\begin{figure}
\centering
\includegraphics[scale=0.45]{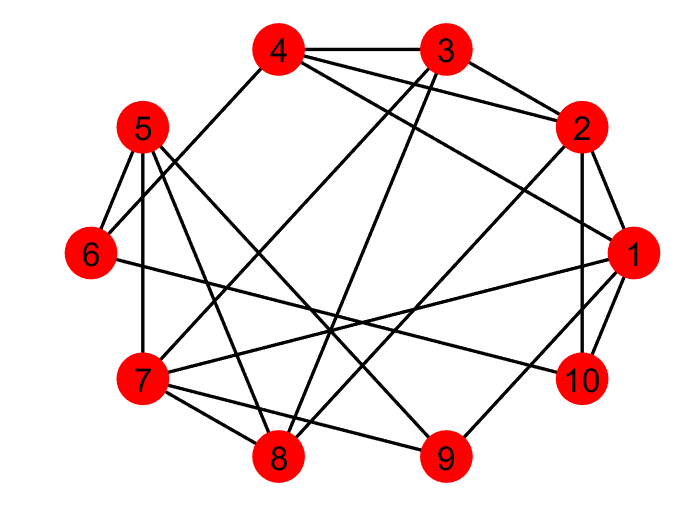}
\caption{Network topology of $N=10$ oscillators.}
\label{fig:N10WS}
\end{figure}
In 1998, Watts and Strogatz proposed that biological or social networks are between completely regular and random, and established the WS model to generate this kind of networks~\cite{watts1998collective}. This model starts from a completely regular network and then randomly reconnects the nodes in the network. As a result, all the path lengths in the network are short and the clustering degree is high, which is consistent with the small-world characteristics~\cite{marconi1909wireless} and carries on rich statistical and dynamics behaviors. So it is interesting to extend application of the Koopman method to systems defined on WS networks.

For ease of observation, we generate a network topology of $N=10$ oscillators in Fig.~\ref{fig:N10WS} , where the ten circled red dots symbol ten oscillators, black numbers marking indices of oscillators, and black edges indicating the interaction topology. The edge connection between vertex $i$ and $j$ gives $a_{ij}=1$ in Eq.~\eqref{equ:KMGeneral}.
\subsection{Koopman analysis}
\begin{figure*}
  \centering
  \subfigure[\quad LD]{ \label{fig:N10LinearDis}
    \begin{minipage}[b]{0.45\textwidth}
    \centering
    \includegraphics[width=1\textwidth]{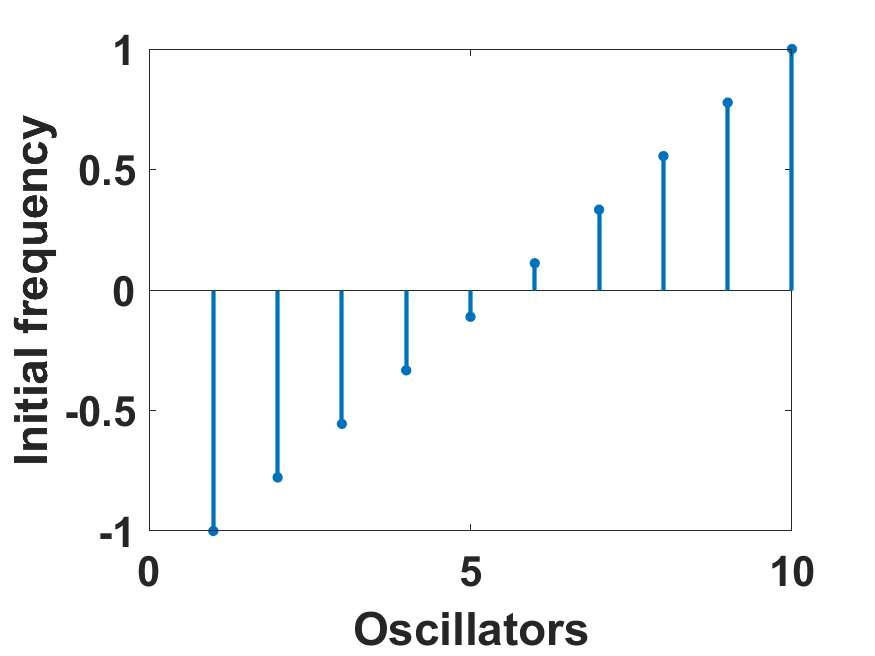}
    \end{minipage}
  }
   \subfigure[\quad RUD]{ \label{fig:N10RUD}
    \begin{minipage}[b]{0.45\textwidth}
    \centering
    \includegraphics[width=1\textwidth]{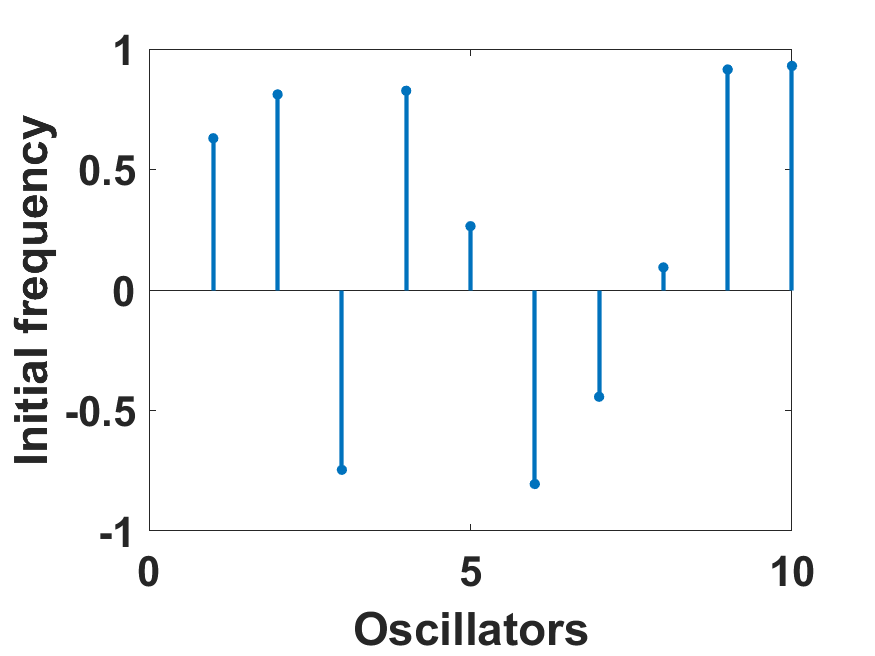}
    \end{minipage}
  }
  \caption{Two frequency distributions for the Kuramoto model on a network. (a) a linear distribution; (b) a realization from the uniform distribution on $[-1,1]$. The network is displayed in Fig.~\ref{fig:N10WS} .}
   \label{fig:InitialFrequency_N10}
\end{figure*}
\begin{figure*}
  \centering
  \subfigure[\quad LD]{ \label{fig:N10BifurcationSub1}
    \begin{minipage}[b]{0.45\textwidth}
    \centering
    \includegraphics[width=1\textwidth]{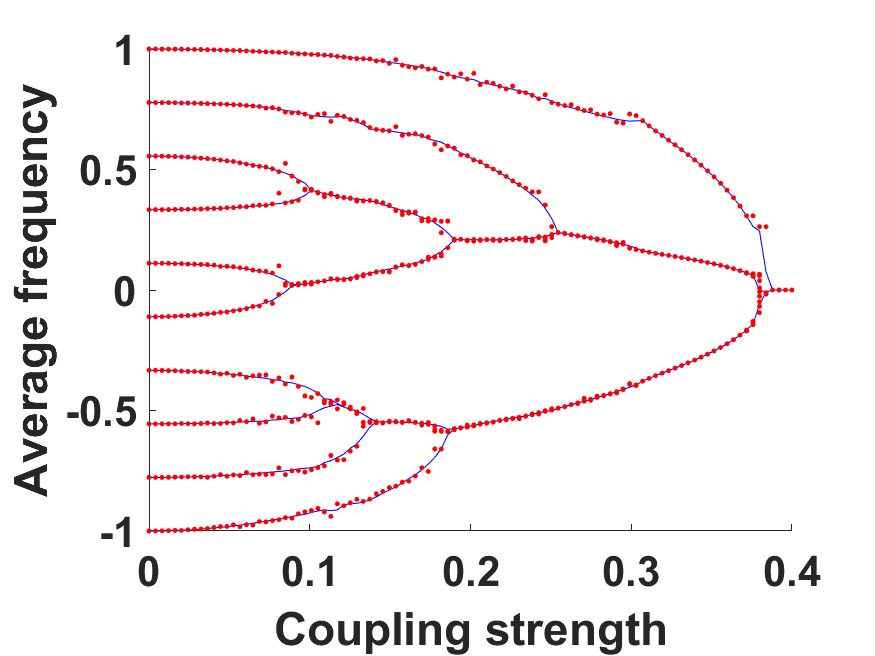}
    \end{minipage}
  }
   \subfigure[\quad RUD]{ \label{fig:N10BifurcationSub2}
    \begin{minipage}[b]{0.45\textwidth}
    \centering
    \includegraphics[width=1\textwidth]{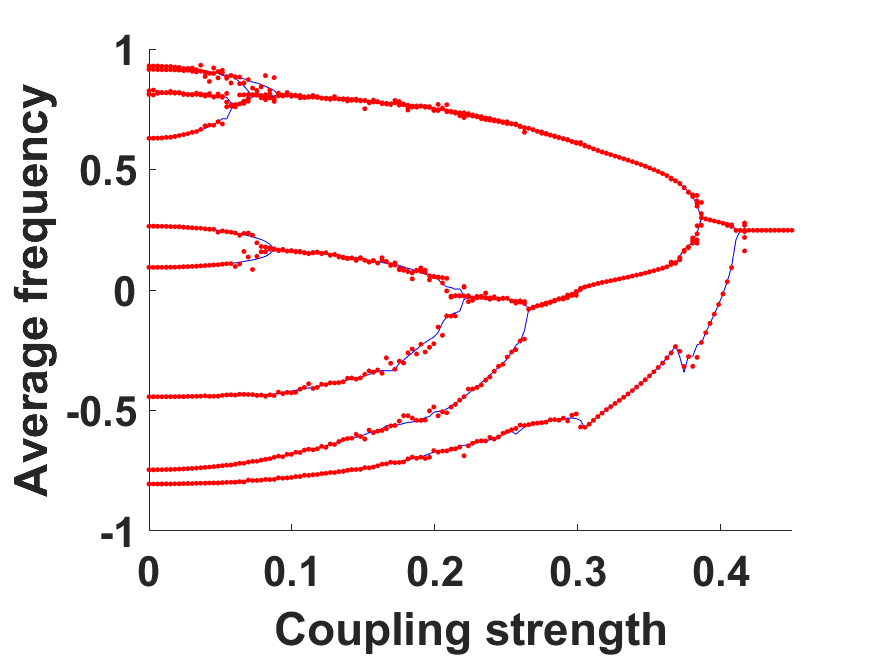}
    \end{minipage}
  }
  \caption{The average frequency {\em vs} the coupling strength $K$ for the Kuramoto model on a network shown in Fig.~\ref{fig:N10WS}, for different generic frequency distributions: (a) linear; (b) random as displayed in Fig.~\ref{fig:InitialFrequency_N10} , and for different computations: the direct numerical average (blue solid lines) and the Koopman analysis (red hollow circles).}
  \label{fig:N10Bifurcation}
\end{figure*}
\begin{figure*}
  \centering
  \subfigure[\quad LD]{\label{fig:correlationCoefficientN10WSTLD}
    \begin{minipage}[b]{0.45\textwidth}
    \centering
    \includegraphics[width=1\textwidth]{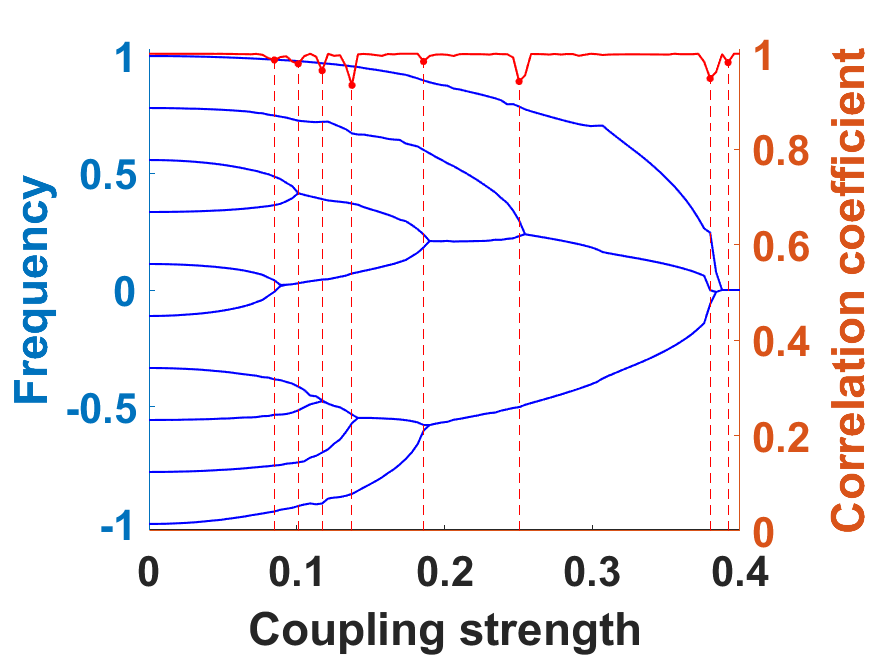}
    \end{minipage}
  }
   \subfigure[\quad RUD]{\label{fig:correlationCoefficientN10WSTRUD}
    \begin{minipage}[b]{0.45\textwidth}
    \centering
    \includegraphics[width=1\textwidth]{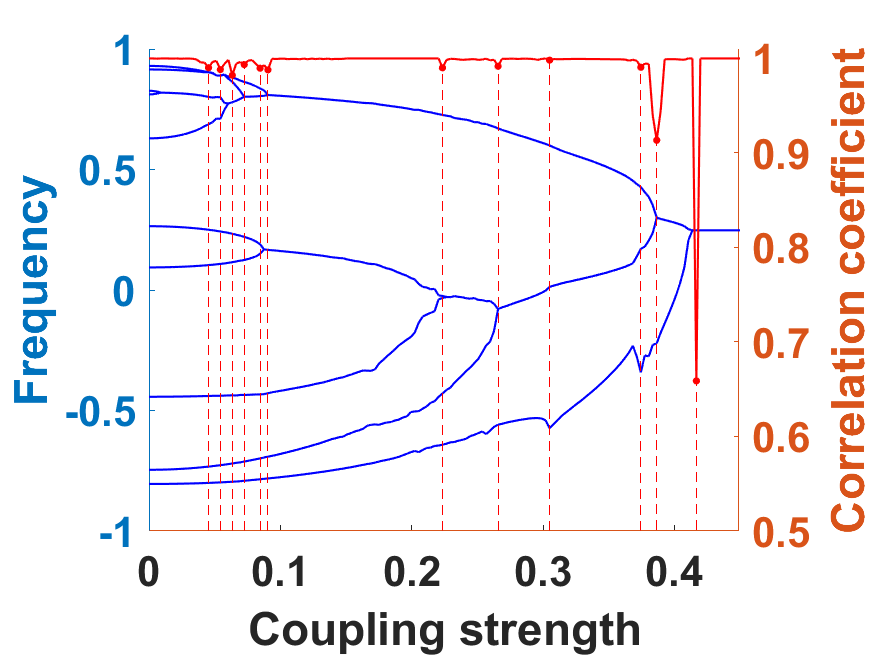}
    \end{minipage}
  }
  \caption{The correlation coefficient (red dashed line) of the eigenfunction and the average frequency (blue solid lines) {\em vs} the coupling strength $K$ . All the parameters are the same as in Fig.~\ref{fig:N10Bifurcation} .}
    \label{fig:correlationCoefficientN10WS}
\end{figure*}
\begin{table*}
\centering
\caption{The critical coupling strength obtained by the direct numerical and the Koopman computation for the Kuramoto model on a complex nextwork shown in Fig.~\ref{fig:N10WS} with the natural frequency linearly distributed in the interval $[-1,1]$ .}
\begin{ruledtabular}
\begin{tabular}{ccccccccc}
numerical &0.088&0.100&0.116&0.140&0.190&0.254&0.384&0.388\\
Koopman   &0.086&0.098&0.116&0.134&0.190&0.246&0.380&0.392\\
\end{tabular}
\end{ruledtabular}
\label{tab:N10SolutionCompareLinear}
\end{table*}
\begin{table*}
\centering
\caption{The critical coupling strength obtained by the direct numerical and the Koopman computation for the Kuramoto model on a complex nextwork shown in Fig.~\ref{fig:N10WS} with the natural frequency from the uniform distribution in the interval $[-1,1]$ .}
\begin{ruledtabular}
\begin{tabular}{ccccccccccc}
numerical &0.009&0.045&0.060&0.072&0.087&0.090&0.222&0.267&0.387&0.414\\
Koopman &-&0.045&0.054,0.063&0.072&0.084&0.090&0.222&0.267&0.387&0.417\\
\end{tabular}
\end{ruledtabular}
\label{tab:N10SolutionCompareRUD}
\end{table*}
With the network topology in Fig.~\ref{fig:N10WS}, here we focus on two typical distributions of frequencies: randomly distributed in the interval $[-1,1]$ with uniform probability (called random-uniform distribution, RUD) and chosen linearly within $[-1,1]$ (called linear distribution, LD) as depicted in Fig.~\ref{fig:InitialFrequency_N10}.

The bifurcation of average frequencies obtained by direct numerical computation is plotted in Fig.~\ref{fig:N10Bifurcation} with blue solid lines. The frequency distribution and coupling strength between oscillators have an obvious impact on the process of synchronization. The oscillators with similar frequencies tend to be synchronized locally first. With the coupling strength increasing, the noncoherent state gets reduced and coherence begins to gain weight gradually. After local synchronization taking place at a larger and larger scale, the global synchronization is finally achieved at certain critical coupling strength. We plot the most important set of frequencies from the Koopman analysis with red hollow circles in Fig.~\ref{fig:N10Bifurcation} . Obviously, the results are in general quite reliable except at local bifurcation points, which may be attributed to the fact that here the interference between oscillators is subtle and sensitive.

We plot the "$\rho$-$K$" graph for this Kuramoto model with red dashed line in Fig.~\ref{fig:correlationCoefficientN10WS} and the natural frequency distribution is LD in Fig.~\ref{fig:correlationCoefficientN10WSTLD} and RUD in Fig.~\ref{fig:correlationCoefficientN10WSTRUD} respectively. The dips in the correlation profile correspond well to phase transition points in the frequency tree. Tab.~\ref{tab:N10SolutionCompareLinear} show the critical coupling strength obtained with direct numerical computation and the frequency jump points identified with the Koopman analysis, where $K_j-K_{j-1}=0.002$. Unfortunately, the phase transition point $K=0.009$ is not detected by the Koopman technique in Tab.~\ref{tab:N10SolutionCompareRUD} with $K_j-K_{j-1}=0.003$. Because the difference of the natural frequency between the two oscillators is tiny and the synchronization occurs at a very small coupling strength (close to zero). In this case, the insignificant change of modes is not able to catch the phase transition point. It is worth mentioning that the phase synchronization detected with direct numerical computation at $K=0.060$ covers two dips in the correlation coefficient at $K=0.054,0.063$ , which may indicate possible occurrence of very subtle bifurcation. The marching stepsize of the coupling strength is smaller than in Sec.~\ref{sec:koopmanForLine}, resulting in a computation error less than 0.01.

In conclusion, the results of the Koopman analysis remains reliable for the Kuramoto model on a complex network.
\section{\label{sec:conclusions}Conclusions}
With the help of spectral properties of Koopman operator, we are able to divide the phase space, extract important dynamics patterns, and determine global properties of coupled nonlinear systems. In this paper, this technique has been extended to the analysis of Kuramoto oscillators on a ring or a complex network. After a Hankel matrix is built from the time series and is preprocessed with SVD, we carry out the DMD based on an approximation of the Koopman operator. Eigenfunctions and eigenvalues are obtained to unfold the frequency structure, and dominant modes are determined by checking the weight of different eigenfunctions. With an RG analysis valid in case of small coupling, the eigen-frequencies turn out to be the average frequencies of the oscillators, which holds true even for moderate or strong couplings justified by the Koopman approach. The correlation of eigenfunctions at neighbouring coupling strength is a manifestation of the change rate of the important dynamical modes, through which local or global synchronization points can be found with small error bars. These new techniques may provide extra tools for the extraction of important dynamical information in complex systems.

In the above analysis, near the bifurcation point, the frequency derived from the Koopman analysis sometimes does not match well with the average frequency. How to improve the computation and pin down important frequencies should be better probed. We use the correlation function as an indicator for the bifurcation events, which seems valid in most cases but may not be the best ones since a majority of other eigenfunctions are ignored. Some important features of dynamics may be lost in the current treatment and so it would be very helpful if one can find a way to pick up contributions from them. Also, in the current investigation, enough data is provided for our purpose. If only a limited amount of data is available, what kind of information can be extracted from the Koopman analysis is an interesting question, which may depend on the functional basis we use to construct the approximate Koopman operator.

Looking back on the whole analysis, the Koopman technique does not require governing equations of motion, and a time series suffices for all the above operations. To reduce the complexity associated with high dimensionality and ease the difficulty of basis selection, we combined the concept of DMD with the time delay embedding and the SVD. Under the premise of ensuring stability and reliability, the Koopman analysis helps us reconstruct global information from partial observation and at the same time minimizes the impact by unavoidable noise. The importance of the average frequency unexpectedly announces again in this type of analysis of complex systems although rhythms in nature appear everywhere and have been encapsulating us already for many years up to this very day. The Koopman analysis seems to provide a new frame for the analysis and understanding of all these fascinating nonlinear systems.
\section*{ACKNOWLEDGMENTS}
This work was supported by the National Natural Science Foundation of China under Grants No.11775035 and also by the Fundamental Research Funds for the Central Universities with Contract No.2019XD-A10.


\begin{thebibliography}{58}%
\makeatletter
\providecommand \@ifxundefined [1]{%
 \@ifx{#1\undefined}
}%
\providecommand \@ifnum [1]{%
 \ifnum #1\expandafter \@firstoftwo
 \else \expandafter \@secondoftwo
 \fi
}%
\providecommand \@ifx [1]{%
 \ifx #1\expandafter \@firstoftwo
 \else \expandafter \@secondoftwo
 \fi
}%
\providecommand \natexlab [1]{#1}%
\providecommand \enquote  [1]{``#1''}%
\providecommand \bibnamefont  [1]{#1}%
\providecommand \bibfnamefont [1]{#1}%
\providecommand \citenamefont [1]{#1}%
\providecommand \href@noop [0]{\@secondoftwo}%
\providecommand \href [0]{\begingroup \@sanitize@url \@href}%
\providecommand \@href[1]{\@@startlink{#1}\@@href}%
\providecommand \@@href[1]{\endgroup#1\@@endlink}%
\providecommand \@sanitize@url [0]{\catcode `\\12\catcode `\$12\catcode
  `\&12\catcode `\#12\catcode `\^12\catcode `\_12\catcode `\%12\relax}%
\providecommand \@@startlink[1]{}%
\providecommand \@@endlink[0]{}%
\providecommand \url  [0]{\begingroup\@sanitize@url \@url }%
\providecommand \@url [1]{\endgroup\@href {#1}{\urlprefix }}%
\providecommand \urlprefix  [0]{URL }%
\providecommand \Eprint [0]{\href }%
\providecommand \doibase [0]{http://dx.doi.org/}%
\providecommand \selectlanguage [0]{\@gobble}%
\providecommand \bibinfo  [0]{\@secondoftwo}%
\providecommand \bibfield  [0]{\@secondoftwo}%
\providecommand \translation [1]{[#1]}%
\providecommand \BibitemOpen [0]{}%
\providecommand \bibitemStop [0]{}%
\providecommand \bibitemNoStop [0]{.\EOS\space}%
\providecommand \EOS [0]{\spacefactor3000\relax}%
\providecommand \BibitemShut  [1]{\csname bibitem#1\endcsname}%
\let\auto@bib@innerbib\@empty
\bibitem [{\citenamefont {Brunton}\ \emph {et~al.}(2017)\citenamefont
  {Brunton}, \citenamefont {Brunton}, \citenamefont {Proctor}, \citenamefont
  {Kaiser},\ and\ \citenamefont {Kutz}}]{brunton2017chaos}%
  \BibitemOpen
  \bibfield  {author} {\bibinfo {author} {\bibfnamefont {S.~L.}\ \bibnamefont
  {Brunton}}, \bibinfo {author} {\bibfnamefont {B.~W.}\ \bibnamefont
  {Brunton}}, \bibinfo {author} {\bibfnamefont {J.~L.}\ \bibnamefont
  {Proctor}}, \bibinfo {author} {\bibfnamefont {E.}~\bibnamefont {Kaiser}}, \
  and\ \bibinfo {author} {\bibfnamefont {J.~N.}\ \bibnamefont {Kutz}},\
  }\href@noop {} {\bibfield  {journal} {\bibinfo  {journal} {Nat. Commun.}\
  }\textbf {\bibinfo {volume} {8}},\ \bibinfo {pages} {1} (\bibinfo {year}
  {2017})}\BibitemShut {NoStop}%
\bibitem [{\citenamefont {Koopman}(1931)}]{koopman1931hamiltonian}%
  \BibitemOpen
  \bibfield  {author} {\bibinfo {author} {\bibfnamefont {B.~O.}\ \bibnamefont
  {Koopman}},\ }\href@noop {} {\bibfield  {journal} {\bibinfo  {journal} {Proc.
  Natl. Acad. Sci. U.S.A.}\ }\textbf {\bibinfo {volume} {17}},\ \bibinfo
  {pages} {315} (\bibinfo {year} {1931})}\BibitemShut {NoStop}%
\bibitem [{\citenamefont {Koopman}\ and\ \citenamefont
  {Neumann}(1932)}]{koopman1932dynamical}%
  \BibitemOpen
  \bibfield  {author} {\bibinfo {author} {\bibfnamefont {B.}~\bibnamefont
  {Koopman}}\ and\ \bibinfo {author} {\bibfnamefont {J.~v.}\ \bibnamefont
  {Neumann}},\ }\href@noop {} {\bibfield  {journal} {\bibinfo  {journal} {Proc.
  Natl. Acad. Sci. U.S.A.}\ }\textbf {\bibinfo {volume} {18}},\ \bibinfo
  {pages} {255} (\bibinfo {year} {1932})}\BibitemShut {NoStop}%
\bibitem [{\citenamefont {Brunton}\ \emph
  {et~al.}(2016{\natexlab{a}})\citenamefont {Brunton}, \citenamefont {Brunton},
  \citenamefont {Proctor},\ and\ \citenamefont {Kutz}}]{brunton2016koopman}%
  \BibitemOpen
  \bibfield  {author} {\bibinfo {author} {\bibfnamefont {S.~L.}\ \bibnamefont
  {Brunton}}, \bibinfo {author} {\bibfnamefont {B.~W.}\ \bibnamefont
  {Brunton}}, \bibinfo {author} {\bibfnamefont {J.~L.}\ \bibnamefont
  {Proctor}}, \ and\ \bibinfo {author} {\bibfnamefont {J.~N.}\ \bibnamefont
  {Kutz}},\ }\href@noop {} {\bibfield  {journal} {\bibinfo  {journal} {PLoS
  One}\ }\textbf {\bibinfo {volume} {11}} (\bibinfo {year}
  {2016}{\natexlab{a}})}\BibitemShut {NoStop}%
\bibitem [{\citenamefont {Arbabi}\ and\ \citenamefont
  {Mezic}(2017)}]{arbabi2017ergodic}%
  \BibitemOpen
  \bibfield  {author} {\bibinfo {author} {\bibfnamefont {H.}~\bibnamefont
  {Arbabi}}\ and\ \bibinfo {author} {\bibfnamefont {I.}~\bibnamefont {Mezic}},\
  }\href@noop {} {\bibfield  {journal} {\bibinfo  {journal} {SIAM J. Appl. Dyn.
  Syst.}\ }\textbf {\bibinfo {volume} {16}},\ \bibinfo {pages} {2096} (\bibinfo
  {year} {2017})}\BibitemShut {NoStop}%
\bibitem [{\citenamefont {Mezi{\'c}}\ and\ \citenamefont
  {Banaszuk}(2004)}]{mezic2004comparison}%
  \BibitemOpen
  \bibfield  {author} {\bibinfo {author} {\bibfnamefont {I.}~\bibnamefont
  {Mezi{\'c}}}\ and\ \bibinfo {author} {\bibfnamefont {A.}~\bibnamefont
  {Banaszuk}},\ }\href@noop {} {\bibfield  {journal} {\bibinfo  {journal}
  {Phys. D}\ }\textbf {\bibinfo {volume} {197}},\ \bibinfo {pages} {101}
  (\bibinfo {year} {2004})}\BibitemShut {NoStop}%
\bibitem [{\citenamefont {Mauroy}\ and\ \citenamefont
  {Mezi{\'c}}(2016)}]{mauroy2016global}%
  \BibitemOpen
  \bibfield  {author} {\bibinfo {author} {\bibfnamefont {A.}~\bibnamefont
  {Mauroy}}\ and\ \bibinfo {author} {\bibfnamefont {I.}~\bibnamefont
  {Mezi{\'c}}},\ }\href@noop {} {\bibfield  {journal} {\bibinfo  {journal}
  {IEEE Trans. Autom. Control}\ }\textbf {\bibinfo {volume} {61}},\ \bibinfo
  {pages} {3356} (\bibinfo {year} {2016})}\BibitemShut {NoStop}%
\bibitem [{\citenamefont {Mezi{\'c}}(2005)}]{mezic2005spectral}%
  \BibitemOpen
  \bibfield  {author} {\bibinfo {author} {\bibfnamefont {I.}~\bibnamefont
  {Mezi{\'c}}},\ }\href@noop {} {\bibfield  {journal} {\bibinfo  {journal}
  {Nonlinear Dyn.}\ }\textbf {\bibinfo {volume} {41}},\ \bibinfo {pages} {309}
  (\bibinfo {year} {2005})}\BibitemShut {NoStop}%
\bibitem [{\citenamefont {Goldenfeld}(2018)}]{goldenfeld2018lectures}%
  \BibitemOpen
  \bibfield  {author} {\bibinfo {author} {\bibfnamefont {N.}~\bibnamefont
  {Goldenfeld}},\ }\href@noop {} {\emph {\bibinfo {title} {Lectures on phase
  transitions and the renormalization group}}}\ (\bibinfo  {publisher} {CRC
  Press},\ \bibinfo {year} {2018})\BibitemShut {NoStop}%
\bibitem [{\citenamefont {Zinn-Justin}(1996)}]{zinn1996quantum}%
  \BibitemOpen
  \bibfield  {author} {\bibinfo {author} {\bibfnamefont {J.}~\bibnamefont
  {Zinn-Justin}},\ }\href@noop {} {\emph {\bibinfo {title} {Quantum field
  theory and critical phenomena}}}\ (\bibinfo  {publisher} {Clarendon Press},\
  \bibinfo {year} {1996})\BibitemShut {NoStop}%
\bibitem [{\citenamefont {Chen}\ \emph
  {et~al.}(1994{\natexlab{a}})\citenamefont {Chen}, \citenamefont {Goldenfeld},
  \citenamefont {Oono},\ and\ \citenamefont {Paquette}}]{chen1994selection}%
  \BibitemOpen
  \bibfield  {author} {\bibinfo {author} {\bibfnamefont {L.-Y.}\ \bibnamefont
  {Chen}}, \bibinfo {author} {\bibfnamefont {N.}~\bibnamefont {Goldenfeld}},
  \bibinfo {author} {\bibfnamefont {Y.}~\bibnamefont {Oono}}, \ and\ \bibinfo
  {author} {\bibfnamefont {G.}~\bibnamefont {Paquette}},\ }\href@noop {}
  {\bibfield  {journal} {\bibinfo  {journal} {Physica A}\ }\textbf {\bibinfo
  {volume} {204}},\ \bibinfo {pages} {111} (\bibinfo {year}
  {1994}{\natexlab{a}})}\BibitemShut {NoStop}%
\bibitem [{\citenamefont {Paquette}\ \emph {et~al.}(1994)\citenamefont
  {Paquette}, \citenamefont {Chen}, \citenamefont {Goldenfeld},\ and\
  \citenamefont {Oono}}]{paquette1994structural}%
  \BibitemOpen
  \bibfield  {author} {\bibinfo {author} {\bibfnamefont {G.}~\bibnamefont
  {Paquette}}, \bibinfo {author} {\bibfnamefont {L.-Y.}\ \bibnamefont {Chen}},
  \bibinfo {author} {\bibfnamefont {N.}~\bibnamefont {Goldenfeld}}, \ and\
  \bibinfo {author} {\bibfnamefont {Y.}~\bibnamefont {Oono}},\ }\href@noop {}
  {\bibfield  {journal} {\bibinfo  {journal} {Phys. Rev. Lett.}\ }\textbf
  {\bibinfo {volume} {72}},\ \bibinfo {pages} {76} (\bibinfo {year}
  {1994})}\BibitemShut {NoStop}%
\bibitem [{\citenamefont {Vakakis}\ and\ \citenamefont
  {Azeez}(1998)}]{vakakis1998analytic}%
  \BibitemOpen
  \bibfield  {author} {\bibinfo {author} {\bibfnamefont {A.~F.}\ \bibnamefont
  {Vakakis}}\ and\ \bibinfo {author} {\bibfnamefont {M.}~\bibnamefont
  {Azeez}},\ }\href@noop {} {\bibfield  {journal} {\bibinfo  {journal}
  {Nonlinear Dyn.}\ }\textbf {\bibinfo {volume} {15}},\ \bibinfo {pages} {245}
  (\bibinfo {year} {1998})}\BibitemShut {NoStop}%
\bibitem [{\citenamefont {Kevorkian}\ \emph {et~al.}(1982)\citenamefont
  {Kevorkian}, \citenamefont {Cole},\ and\ \citenamefont
  {Nayfeh}}]{kevorkian1982perturbation}%
  \BibitemOpen
  \bibfield  {author} {\bibinfo {author} {\bibfnamefont {J.}~\bibnamefont
  {Kevorkian}}, \bibinfo {author} {\bibfnamefont {J.}~\bibnamefont {Cole}}, \
  and\ \bibinfo {author} {\bibfnamefont {A.~H.}\ \bibnamefont {Nayfeh}},\
  }\href@noop {} {\bibfield  {journal} {\bibinfo  {journal} {Bulletin of the
  American Mathematical Society}\ }\textbf {\bibinfo {volume} {7}},\ \bibinfo
  {pages} {414} (\bibinfo {year} {1982})}\BibitemShut {NoStop}%
\bibitem [{\citenamefont {Bender}\ and\ \citenamefont
  {Orszag}(2013)}]{bender2013advanced}%
  \BibitemOpen
  \bibfield  {author} {\bibinfo {author} {\bibfnamefont {C.~M.}\ \bibnamefont
  {Bender}}\ and\ \bibinfo {author} {\bibfnamefont {S.~A.}\ \bibnamefont
  {Orszag}},\ }\href@noop {} {\emph {\bibinfo {title} {Advanced mathematical
  methods for scientists and engineers I: Asymptotic methods and perturbation
  theory}}}\ (\bibinfo  {publisher} {Springer Science \& Business Media},\
  \bibinfo {year} {2013})\BibitemShut {NoStop}%
\bibitem [{\citenamefont {Lan}(2013)}]{lan2013bridging}%
  \BibitemOpen
  \bibfield  {author} {\bibinfo {author} {\bibfnamefont {Y.}~\bibnamefont
  {Lan}},\ }\href@noop {} {\bibfield  {journal} {\bibinfo  {journal} {Phys.
  Rev. E}\ }\textbf {\bibinfo {volume} {87}},\ \bibinfo {pages} {012914}
  (\bibinfo {year} {2013})}\BibitemShut {NoStop}%
\bibitem [{\citenamefont {Zheng}\ \emph {et~al.}(2012)\citenamefont {Zheng},
  \citenamefont {Hu},\ and\ \citenamefont {Hu}}]{Zheng2012Phase}%
  \BibitemOpen
  \bibfield  {author} {\bibinfo {author} {\bibfnamefont {Z.}~\bibnamefont
  {Zheng}}, \bibinfo {author} {\bibfnamefont {G.}~\bibnamefont {Hu}}, \ and\
  \bibinfo {author} {\bibfnamefont {B.}~\bibnamefont {Hu}},\ }\href@noop {}
  {\bibfield  {journal} {\bibinfo  {journal} {Phys. Rev. Lett.}\ }\textbf
  {\bibinfo {volume} {81}},\ \bibinfo {pages} {5318} (\bibinfo {year}
  {2012})}\BibitemShut {NoStop}%
\bibitem [{\citenamefont {Rosenblum}\ \emph {et~al.}(1996)\citenamefont
  {Rosenblum}, \citenamefont {Pikovsky},\ and\ \citenamefont
  {Kurths}}]{Rosenblum1996Phase}%
  \BibitemOpen
  \bibfield  {author} {\bibinfo {author} {\bibfnamefont {M.~G.}\ \bibnamefont
  {Rosenblum}}, \bibinfo {author} {\bibfnamefont {A.~S.}\ \bibnamefont
  {Pikovsky}}, \ and\ \bibinfo {author} {\bibfnamefont {J.}~\bibnamefont
  {Kurths}},\ }\href@noop {} {\bibfield  {journal} {\bibinfo  {journal} {Phys.
  Rev. Lett.}\ }\textbf {\bibinfo {volume} {76}},\ \bibinfo {pages} {1804}
  (\bibinfo {year} {1996})}\BibitemShut {NoStop}%
\bibitem [{\citenamefont {Acebr{\'o}n}\ \emph {et~al.}(2005)\citenamefont
  {Acebr{\'o}n}, \citenamefont {Bonilla}, \citenamefont {Vicente},
  \citenamefont {Ritort},\ and\ \citenamefont {Spigler}}]{acebron2005kuramoto}%
  \BibitemOpen
  \bibfield  {author} {\bibinfo {author} {\bibfnamefont {J.~A.}\ \bibnamefont
  {Acebr{\'o}n}}, \bibinfo {author} {\bibfnamefont {L.~L.}\ \bibnamefont
  {Bonilla}}, \bibinfo {author} {\bibfnamefont {C.~J.~P.}\ \bibnamefont
  {Vicente}}, \bibinfo {author} {\bibfnamefont {F.}~\bibnamefont {Ritort}}, \
  and\ \bibinfo {author} {\bibfnamefont {R.}~\bibnamefont {Spigler}},\
  }\href@noop {} {\bibfield  {journal} {\bibinfo  {journal} {Rev. Mod. Phys.}\
  }\textbf {\bibinfo {volume} {77}},\ \bibinfo {pages} {137} (\bibinfo {year}
  {2005})}\BibitemShut {NoStop}%
\bibitem [{\citenamefont {Chiba}\ \emph {et~al.}(2018)\citenamefont {Chiba},
  \citenamefont {Medvedev},\ and\ \citenamefont
  {Mizuhara}}]{chiba2018bifurcations}%
  \BibitemOpen
  \bibfield  {author} {\bibinfo {author} {\bibfnamefont {H.}~\bibnamefont
  {Chiba}}, \bibinfo {author} {\bibfnamefont {G.~S.}\ \bibnamefont {Medvedev}},
  \ and\ \bibinfo {author} {\bibfnamefont {M.~S.}\ \bibnamefont {Mizuhara}},\
  }\href@noop {} {\bibfield  {journal} {\bibinfo  {journal} {Chaos: An
  Interdisciplinary Journal of Nonlinear Science}\ }\textbf {\bibinfo {volume}
  {28}},\ \bibinfo {pages} {073109} (\bibinfo {year} {2018})}\BibitemShut
  {NoStop}%
\bibitem [{\citenamefont {Wu}\ \emph {et~al.}(2012)\citenamefont {Wu},
  \citenamefont {Xiao}, \citenamefont {Hu},\ and\ \citenamefont
  {Zhan}}]{Wu2012Synchronizing}%
  \BibitemOpen
  \bibfield  {author} {\bibinfo {author} {\bibfnamefont {Y.}~\bibnamefont
  {Wu}}, \bibinfo {author} {\bibfnamefont {J.}~\bibnamefont {Xiao}}, \bibinfo
  {author} {\bibfnamefont {G.}~\bibnamefont {Hu}}, \ and\ \bibinfo {author}
  {\bibfnamefont {M.}~\bibnamefont {Zhan}},\ }\href@noop {} {\bibfield
  {journal} {\bibinfo  {journal} {EPL (Europhysics Letters)}\ }\textbf
  {\bibinfo {volume} {97}},\ \bibinfo {pages} {40005} (\bibinfo {year}
  {2012})}\BibitemShut {NoStop}%
\bibitem [{\citenamefont {Strogatz}(2000)}]{strogatz2000kuramoto}%
  \BibitemOpen
  \bibfield  {author} {\bibinfo {author} {\bibfnamefont {S.~H.}\ \bibnamefont
  {Strogatz}},\ }\href@noop {} {\bibfield  {journal} {\bibinfo  {journal}
  {Phys. D}\ }\textbf {\bibinfo {volume} {143}},\ \bibinfo {pages} {1}
  (\bibinfo {year} {2000})}\BibitemShut {NoStop}%
\bibitem [{\citenamefont {Ott}\ and\ \citenamefont
  {Antonsen}(2008)}]{ott2008low}%
  \BibitemOpen
  \bibfield  {author} {\bibinfo {author} {\bibfnamefont {E.}~\bibnamefont
  {Ott}}\ and\ \bibinfo {author} {\bibfnamefont {T.~M.}\ \bibnamefont
  {Antonsen}},\ }\href@noop {} {\bibfield  {journal} {\bibinfo  {journal}
  {Chaos: An Interdisciplinary Journal of Nonlinear Science}\ }\textbf
  {\bibinfo {volume} {18}},\ \bibinfo {pages} {037113} (\bibinfo {year}
  {2008})}\BibitemShut {NoStop}%
\bibitem [{\citenamefont {Ott}\ and\ \citenamefont
  {Antonsen}(2009)}]{ott2009long}%
  \BibitemOpen
  \bibfield  {author} {\bibinfo {author} {\bibfnamefont {E.}~\bibnamefont
  {Ott}}\ and\ \bibinfo {author} {\bibfnamefont {T.~M.}\ \bibnamefont
  {Antonsen}},\ }\href@noop {} {\bibfield  {journal} {\bibinfo  {journal}
  {Chaos: An interdisciplinary journal of nonlinear science}\ }\textbf
  {\bibinfo {volume} {19}},\ \bibinfo {pages} {023117} (\bibinfo {year}
  {2009})}\BibitemShut {NoStop}%
\bibitem [{\citenamefont {Martens}\ \emph {et~al.}(2009)\citenamefont
  {Martens}, \citenamefont {Barreto}, \citenamefont {Strogatz}, \citenamefont
  {Ott}, \citenamefont {So},\ and\ \citenamefont
  {Antonsen}}]{martens2009exact}%
  \BibitemOpen
  \bibfield  {author} {\bibinfo {author} {\bibfnamefont {E.~A.}\ \bibnamefont
  {Martens}}, \bibinfo {author} {\bibfnamefont {E.}~\bibnamefont {Barreto}},
  \bibinfo {author} {\bibfnamefont {S.~H.}\ \bibnamefont {Strogatz}}, \bibinfo
  {author} {\bibfnamefont {E.}~\bibnamefont {Ott}}, \bibinfo {author}
  {\bibfnamefont {P.}~\bibnamefont {So}}, \ and\ \bibinfo {author}
  {\bibfnamefont {T.~M.}\ \bibnamefont {Antonsen}},\ }\href@noop {} {\bibfield
  {journal} {\bibinfo  {journal} {Phys. Rev. E}\ }\textbf {\bibinfo {volume}
  {79}},\ \bibinfo {pages} {026204} (\bibinfo {year} {2009})}\BibitemShut
  {NoStop}%
\bibitem [{\citenamefont {Wu}\ \emph {et~al.}(2016)\citenamefont {Wu},
  \citenamefont {Cheng}, \citenamefont {Dai},\ and\ \citenamefont
  {Li}}]{wu2016ott}%
  \BibitemOpen
  \bibfield  {author} {\bibinfo {author} {\bibfnamefont {N.-P.}\ \bibnamefont
  {Wu}}, \bibinfo {author} {\bibfnamefont {H.-Y.}\ \bibnamefont {Cheng}},
  \bibinfo {author} {\bibfnamefont {Q.-L.}\ \bibnamefont {Dai}}, \ and\
  \bibinfo {author} {\bibfnamefont {H.-H.}\ \bibnamefont {Li}},\ }\href@noop {}
  {\bibfield  {journal} {\bibinfo  {journal} {Chin. Phys. Lett.}\ }\textbf
  {\bibinfo {volume} {33}},\ \bibinfo {pages} {070501} (\bibinfo {year}
  {2016})}\BibitemShut {NoStop}%
\bibitem [{\citenamefont {Moreira}\ and\ \citenamefont
  {de~Aguiar}(2019)}]{moreira2019global}%
  \BibitemOpen
  \bibfield  {author} {\bibinfo {author} {\bibfnamefont {C.~A.}\ \bibnamefont
  {Moreira}}\ and\ \bibinfo {author} {\bibfnamefont {M.~A.}\ \bibnamefont
  {de~Aguiar}},\ }\href@noop {} {\bibfield  {journal} {\bibinfo  {journal}
  {Physica A}\ }\textbf {\bibinfo {volume} {514}},\ \bibinfo {pages} {487}
  (\bibinfo {year} {2019})}\BibitemShut {NoStop}%
\bibitem [{\citenamefont {Zhang}\ and\ \citenamefont
  {Zhu}(2019)}]{zhang2019exponential}%
  \BibitemOpen
  \bibfield  {author} {\bibinfo {author} {\bibfnamefont {J.}~\bibnamefont
  {Zhang}}\ and\ \bibinfo {author} {\bibfnamefont {J.}~\bibnamefont {Zhu}},\
  }\href@noop {} {\bibfield  {journal} {\bibinfo  {journal} {Automatica}\
  }\textbf {\bibinfo {volume} {102}},\ \bibinfo {pages} {122} (\bibinfo {year}
  {2019})}\BibitemShut {NoStop}%
\bibitem [{\citenamefont {Budi{\v{s}}i{\'c}}\ and\ \citenamefont
  {Mezi{\'c}}(2012)}]{budivsic2012geometry}%
  \BibitemOpen
  \bibfield  {author} {\bibinfo {author} {\bibfnamefont {M.}~\bibnamefont
  {Budi{\v{s}}i{\'c}}}\ and\ \bibinfo {author} {\bibfnamefont {I.}~\bibnamefont
  {Mezi{\'c}}},\ }\href@noop {} {\bibfield  {journal} {\bibinfo  {journal}
  {Phys. D}\ }\textbf {\bibinfo {volume} {241}},\ \bibinfo {pages} {1255}
  (\bibinfo {year} {2012})}\BibitemShut {NoStop}%
\bibitem [{\citenamefont {Mezi{\'c}}\ and\ \citenamefont
  {Wiggins}(1999)}]{mezic1999method}%
  \BibitemOpen
  \bibfield  {author} {\bibinfo {author} {\bibfnamefont {I.}~\bibnamefont
  {Mezi{\'c}}}\ and\ \bibinfo {author} {\bibfnamefont {S.}~\bibnamefont
  {Wiggins}},\ }\href@noop {} {\bibfield  {journal} {\bibinfo  {journal}
  {Chaos: An Interdisciplinary Journal of Nonlinear Science}\ }\textbf
  {\bibinfo {volume} {9}},\ \bibinfo {pages} {213} (\bibinfo {year}
  {1999})}\BibitemShut {NoStop}%
\bibitem [{\citenamefont {Budi{\v{s}}i{\'c}}\ and\ \citenamefont
  {Mezi{\'c}}(2009)}]{budivsic2009approximate}%
  \BibitemOpen
  \bibfield  {author} {\bibinfo {author} {\bibfnamefont {M.}~\bibnamefont
  {Budi{\v{s}}i{\'c}}}\ and\ \bibinfo {author} {\bibfnamefont {I.}~\bibnamefont
  {Mezi{\'c}}},\ }in\ \href@noop {} {\emph {\bibinfo {booktitle} {Proceedings
  of the 48h IEEE Conference on Decision and Control (CDC) held jointly with
  2009 28th Chinese Control Conference}}}\ (\bibinfo {organization} {IEEE},\
  \bibinfo {year} {2009})\ pp.\ \bibinfo {pages} {3162--3168}\BibitemShut
  {NoStop}%
\bibitem [{\citenamefont {Mezi{\'c}}(2015)}]{mezic2015applications}%
  \BibitemOpen
  \bibfield  {author} {\bibinfo {author} {\bibfnamefont {I.}~\bibnamefont
  {Mezi{\'c}}},\ }in\ \href@noop {} {\emph {\bibinfo {booktitle} {2015 54th
  IEEE Conference on Decision and Control (CDC)}}}\ (\bibinfo {organization}
  {IEEE},\ \bibinfo {year} {2015})\ pp.\ \bibinfo {pages}
  {7034--7041}\BibitemShut {NoStop}%
\bibitem [{\citenamefont {Kaiser}\ \emph {et~al.}(2017)\citenamefont {Kaiser},
  \citenamefont {Kutz},\ and\ \citenamefont {Brunton}}]{Kaiser2017Sparse}%
  \BibitemOpen
  \bibfield  {author} {\bibinfo {author} {\bibfnamefont {E.}~\bibnamefont
  {Kaiser}}, \bibinfo {author} {\bibfnamefont {J.~N.}\ \bibnamefont {Kutz}}, \
  and\ \bibinfo {author} {\bibfnamefont {S.~L.}\ \bibnamefont {Brunton}},\
  }\href@noop {} {\bibfield  {journal} {\bibinfo  {journal} {Proceedings of the
  Royal Society A Mathematical Physical, and Engineering Sciences}\ }\textbf
  {\bibinfo {volume} {474}} (\bibinfo {year} {2017})}\BibitemShut {NoStop}%
\bibitem [{\citenamefont {Rowley}\ \emph {et~al.}(2009)\citenamefont {Rowley},
  \citenamefont {Mezi{\'c}}, \citenamefont {Bagheri}, \citenamefont
  {Schlatter},\ and\ \citenamefont {Henningson}}]{rowley2009spectral}%
  \BibitemOpen
  \bibfield  {author} {\bibinfo {author} {\bibfnamefont {C.~W.}\ \bibnamefont
  {Rowley}}, \bibinfo {author} {\bibfnamefont {I.}~\bibnamefont {Mezi{\'c}}},
  \bibinfo {author} {\bibfnamefont {S.}~\bibnamefont {Bagheri}}, \bibinfo
  {author} {\bibfnamefont {P.}~\bibnamefont {Schlatter}}, \ and\ \bibinfo
  {author} {\bibfnamefont {D.~S.}\ \bibnamefont {Henningson}},\ }\href@noop {}
  {\bibfield  {journal} {\bibinfo  {journal} {J. Fluid Mech.}\ }\textbf
  {\bibinfo {volume} {641}},\ \bibinfo {pages} {115} (\bibinfo {year}
  {2009})}\BibitemShut {NoStop}%
\bibitem [{\citenamefont {Mezi}\ and\ \citenamefont
  {Igor}(2013)}]{Mezi2013Analysis}%
  \BibitemOpen
  \bibfield  {author} {\bibinfo {author} {\bibnamefont {Mezi}}\ and\ \bibinfo
  {author} {\bibnamefont {Igor}},\ }\href@noop {} {\bibfield  {journal}
  {\bibinfo  {journal} {Annu. Rev. Fluid Mech.}\ }\textbf {\bibinfo {volume}
  {45}},\ \bibinfo {pages} {357} (\bibinfo {year} {2013})}\BibitemShut
  {NoStop}%
\bibitem [{\citenamefont {Brunton}\ \emph
  {et~al.}(2016{\natexlab{b}})\citenamefont {Brunton}, \citenamefont {Johnson},
  \citenamefont {Ojemann},\ and\ \citenamefont {Kutz}}]{brunton2016extracting}%
  \BibitemOpen
  \bibfield  {author} {\bibinfo {author} {\bibfnamefont {B.~W.}\ \bibnamefont
  {Brunton}}, \bibinfo {author} {\bibfnamefont {L.~A.}\ \bibnamefont
  {Johnson}}, \bibinfo {author} {\bibfnamefont {J.~G.}\ \bibnamefont
  {Ojemann}}, \ and\ \bibinfo {author} {\bibfnamefont {J.~N.}\ \bibnamefont
  {Kutz}},\ }\href@noop {} {\bibfield  {journal} {\bibinfo  {journal} {J.
  Neurosci. Methods}\ }\textbf {\bibinfo {volume} {258}},\ \bibinfo {pages} {1}
  (\bibinfo {year} {2016}{\natexlab{b}})}\BibitemShut {NoStop}%
\bibitem [{\citenamefont {Schmid}(2010)}]{schmid2010dynamic}%
  \BibitemOpen
  \bibfield  {author} {\bibinfo {author} {\bibfnamefont {P.~J.}\ \bibnamefont
  {Schmid}},\ }\href@noop {} {\bibfield  {journal} {\bibinfo  {journal} {J.
  Fluid Mech.}\ }\textbf {\bibinfo {volume} {656}},\ \bibinfo {pages} {5}
  (\bibinfo {year} {2010})}\BibitemShut {NoStop}%
\bibitem [{\citenamefont {Sauer}\ \emph {et~al.}(1991)\citenamefont {Sauer},
  \citenamefont {Yorke},\ and\ \citenamefont {Casdagli}}]{sauer1991embedology}%
  \BibitemOpen
  \bibfield  {author} {\bibinfo {author} {\bibfnamefont {T.}~\bibnamefont
  {Sauer}}, \bibinfo {author} {\bibfnamefont {J.~A.}\ \bibnamefont {Yorke}}, \
  and\ \bibinfo {author} {\bibfnamefont {M.}~\bibnamefont {Casdagli}},\
  }\href@noop {} {\bibfield  {journal} {\bibinfo  {journal} {J. Stat. Phys.}\
  }\textbf {\bibinfo {volume} {65}},\ \bibinfo {pages} {579} (\bibinfo {year}
  {1991})}\BibitemShut {NoStop}%
\bibitem [{\citenamefont {Takens}(1981)}]{takens1981detecting}%
  \BibitemOpen
  \bibfield  {author} {\bibinfo {author} {\bibfnamefont {F.}~\bibnamefont
  {Takens}},\ }in\ \href@noop {} {\emph {\bibinfo {booktitle} {Dynamical
  systems and turbulence, Warwick 1980}}}\ (\bibinfo  {publisher} {Springer},\
  \bibinfo {year} {1981})\ pp.\ \bibinfo {pages} {366--381}\BibitemShut
  {NoStop}%
\bibitem [{\citenamefont {Rowley}(2005)}]{rowley2005model}%
  \BibitemOpen
  \bibfield  {author} {\bibinfo {author} {\bibfnamefont {C.~W.}\ \bibnamefont
  {Rowley}},\ }\href@noop {} {\bibfield  {journal} {\bibinfo  {journal} {Int.
  J. Bifurcation Chaos}\ }\textbf {\bibinfo {volume} {15}},\ \bibinfo {pages}
  {997} (\bibinfo {year} {2005})}\BibitemShut {NoStop}%
\bibitem [{\citenamefont {Pikovsky}\ \emph {et~al.}(2003)\citenamefont
  {Pikovsky}, \citenamefont {Kurths}, \citenamefont {Rosenblum},\ and\
  \citenamefont {Kurths}}]{pikovsky2003synchronization}%
  \BibitemOpen
  \bibfield  {author} {\bibinfo {author} {\bibfnamefont {A.}~\bibnamefont
  {Pikovsky}}, \bibinfo {author} {\bibfnamefont {J.}~\bibnamefont {Kurths}},
  \bibinfo {author} {\bibfnamefont {M.}~\bibnamefont {Rosenblum}}, \ and\
  \bibinfo {author} {\bibfnamefont {J.}~\bibnamefont {Kurths}},\ }\href@noop {}
  {\emph {\bibinfo {title} {Synchronization: a universal concept in nonlinear
  sciences}}},\ Vol.~\bibinfo {volume} {12}\ (\bibinfo  {publisher} {Cambridge
  university press},\ \bibinfo {year} {2003})\BibitemShut {NoStop}%
\bibitem [{\citenamefont {Winfree}(1967)}]{winfree1967biological}%
  \BibitemOpen
  \bibfield  {author} {\bibinfo {author} {\bibfnamefont {A.~T.}\ \bibnamefont
  {Winfree}},\ }\href@noop {} {\bibfield  {journal} {\bibinfo  {journal} {J.
  Theor. Biol.}\ }\textbf {\bibinfo {volume} {16}},\ \bibinfo {pages} {15}
  (\bibinfo {year} {1967})}\BibitemShut {NoStop}%
\bibitem [{\citenamefont {Kuramoto}(1975)}]{kuramoto1975international}%
  \BibitemOpen
  \bibfield  {author} {\bibinfo {author} {\bibfnamefont {Y.}~\bibnamefont
  {Kuramoto}},\ }\href@noop {} {\bibfield  {journal} {\bibinfo  {journal}
  {Lect. Notes Phys.}\ }\textbf {\bibinfo {volume} {30}},\ \bibinfo {pages}
  {420} (\bibinfo {year} {1975})}\BibitemShut {NoStop}%
\bibitem [{\citenamefont {Zheng}\ \emph {et~al.}(1998)\citenamefont {Zheng},
  \citenamefont {Hu},\ and\ \citenamefont {Hu}}]{zheng1998phase}%
  \BibitemOpen
  \bibfield  {author} {\bibinfo {author} {\bibfnamefont {Z.}~\bibnamefont
  {Zheng}}, \bibinfo {author} {\bibfnamefont {G.}~\bibnamefont {Hu}}, \ and\
  \bibinfo {author} {\bibfnamefont {B.}~\bibnamefont {Hu}},\ }\href@noop {}
  {\bibfield  {journal} {\bibinfo  {journal} {Phys. Rev. Lett.}\ }\textbf
  {\bibinfo {volume} {81}},\ \bibinfo {pages} {5318} (\bibinfo {year}
  {1998})}\BibitemShut {NoStop}%
\bibitem [{\citenamefont {Kogan}\ \emph {et~al.}(2009)\citenamefont {Kogan},
  \citenamefont {Rogers}, \citenamefont {Cross},\ and\ \citenamefont
  {Refael}}]{kogan2009renormalization}%
  \BibitemOpen
  \bibfield  {author} {\bibinfo {author} {\bibfnamefont {O.}~\bibnamefont
  {Kogan}}, \bibinfo {author} {\bibfnamefont {J.~L.}\ \bibnamefont {Rogers}},
  \bibinfo {author} {\bibfnamefont {M.}~\bibnamefont {Cross}}, \ and\ \bibinfo
  {author} {\bibfnamefont {G.}~\bibnamefont {Refael}},\ }\href@noop {}
  {\bibfield  {journal} {\bibinfo  {journal} {Phys. Rev. E}\ }\textbf {\bibinfo
  {volume} {80}},\ \bibinfo {pages} {036206} (\bibinfo {year}
  {2009})}\BibitemShut {NoStop}%
\bibitem [{\citenamefont {El-Nashar}\ \emph {et~al.}(2009)\citenamefont
  {El-Nashar}, \citenamefont {Muruganandam}, \citenamefont {Ferreira},\ and\
  \citenamefont {Cerdeira}}]{el2009transition}%
  \BibitemOpen
  \bibfield  {author} {\bibinfo {author} {\bibfnamefont {H.~F.}\ \bibnamefont
  {El-Nashar}}, \bibinfo {author} {\bibfnamefont {P.}~\bibnamefont
  {Muruganandam}}, \bibinfo {author} {\bibfnamefont {F.~F.}\ \bibnamefont
  {Ferreira}}, \ and\ \bibinfo {author} {\bibfnamefont {H.~A.}\ \bibnamefont
  {Cerdeira}},\ }\href@noop {} {\bibfield  {journal} {\bibinfo  {journal}
  {Chaos: An Interdisciplinary Journal of Nonlinear Science}\ }\textbf
  {\bibinfo {volume} {19}},\ \bibinfo {pages} {013103} (\bibinfo {year}
  {2009})}\BibitemShut {NoStop}%
\bibitem [{\citenamefont {Chen}\ \emph
  {et~al.}(1994{\natexlab{b}})\citenamefont {Chen}, \citenamefont
  {Goldenfeld},\ and\ \citenamefont {Oono}}]{chen1994renormalization}%
  \BibitemOpen
  \bibfield  {author} {\bibinfo {author} {\bibfnamefont {L.~Y.}\ \bibnamefont
  {Chen}}, \bibinfo {author} {\bibfnamefont {N.}~\bibnamefont {Goldenfeld}}, \
  and\ \bibinfo {author} {\bibfnamefont {Y.}~\bibnamefont {Oono}},\ }\href@noop
  {} {\bibfield  {journal} {\bibinfo  {journal} {Phys. Rev. Lett.}\ }\textbf
  {\bibinfo {volume} {73}},\ \bibinfo {pages} {1311} (\bibinfo {year}
  {1994}{\natexlab{b}})}\BibitemShut {NoStop}%
\bibitem [{\citenamefont {Kunihiro}(1995)}]{kunihiro1995geometrical}%
  \BibitemOpen
  \bibfield  {author} {\bibinfo {author} {\bibfnamefont {T.}~\bibnamefont
  {Kunihiro}},\ }\href@noop {} {\bibfield  {journal} {\bibinfo  {journal}
  {Progress of theoretical physics}\ }\textbf {\bibinfo {volume} {94}},\
  \bibinfo {pages} {503} (\bibinfo {year} {1995})}\BibitemShut {NoStop}%
\bibitem [{\citenamefont {Kunihiro}(1997)}]{kunihiro1997geometrical}%
  \BibitemOpen
  \bibfield  {author} {\bibinfo {author} {\bibfnamefont {T.}~\bibnamefont
  {Kunihiro}},\ }\href@noop {} {\bibfield  {journal} {\bibinfo  {journal} {Jpn.
  J. Ind. Appl. Math.}\ }\textbf {\bibinfo {volume} {14}},\ \bibinfo {pages}
  {51} (\bibinfo {year} {1997})}\BibitemShut {NoStop}%
\bibitem [{\citenamefont {Ei}\ \emph {et~al.}(2000)\citenamefont {Ei},
  \citenamefont {Fujii},\ and\ \citenamefont
  {Kunihiro}}]{ei2000renormalization}%
  \BibitemOpen
  \bibfield  {author} {\bibinfo {author} {\bibfnamefont {S.-I.}\ \bibnamefont
  {Ei}}, \bibinfo {author} {\bibfnamefont {K.}~\bibnamefont {Fujii}}, \ and\
  \bibinfo {author} {\bibfnamefont {T.}~\bibnamefont {Kunihiro}},\ }\href@noop
  {} {\bibfield  {journal} {\bibinfo  {journal} {Ann. Phys.}\ }\textbf
  {\bibinfo {volume} {280}},\ \bibinfo {pages} {236} (\bibinfo {year}
  {2000})}\BibitemShut {NoStop}%
\bibitem [{\citenamefont {Hatta}\ and\ \citenamefont
  {Kunihiro}(2002)}]{hatta2002renormalization}%
  \BibitemOpen
  \bibfield  {author} {\bibinfo {author} {\bibfnamefont {Y.}~\bibnamefont
  {Hatta}}\ and\ \bibinfo {author} {\bibfnamefont {T.}~\bibnamefont
  {Kunihiro}},\ }\href@noop {} {\bibfield  {journal} {\bibinfo  {journal} {Ann.
  Phys.}\ }\textbf {\bibinfo {volume} {298}},\ \bibinfo {pages} {24} (\bibinfo
  {year} {2002})}\BibitemShut {NoStop}%
\bibitem [{\citenamefont
  {Kunihiro}(1998{\natexlab{a}})}]{kunihiro1998renormalization1}%
  \BibitemOpen
  \bibfield  {author} {\bibinfo {author} {\bibfnamefont {T.}~\bibnamefont
  {Kunihiro}},\ }\href@noop {} {\bibfield  {journal} {\bibinfo  {journal}
  {Progress of Theoretical Physics Supplement}\ }\textbf {\bibinfo {volume}
  {131}},\ \bibinfo {pages} {459} (\bibinfo {year}
  {1998}{\natexlab{a}})}\BibitemShut {NoStop}%
\bibitem [{\citenamefont
  {Kunihiro}(1998{\natexlab{b}})}]{kunihiro1998renormalization2}%
  \BibitemOpen
  \bibfield  {author} {\bibinfo {author} {\bibfnamefont {T.}~\bibnamefont
  {Kunihiro}},\ }\href@noop {} {\bibfield  {journal} {\bibinfo  {journal}
  {Phys. Rev. D}\ }\textbf {\bibinfo {volume} {57}},\ \bibinfo {pages} {R2035}
  (\bibinfo {year} {1998}{\natexlab{b}})}\BibitemShut {NoStop}%
\bibitem [{\citenamefont {Kunihiro}\ and\ \citenamefont
  {Matsukidaira}(1998)}]{kunihiro1998dynamical}%
  \BibitemOpen
  \bibfield  {author} {\bibinfo {author} {\bibfnamefont {T.}~\bibnamefont
  {Kunihiro}}\ and\ \bibinfo {author} {\bibfnamefont {J.}~\bibnamefont
  {Matsukidaira}},\ }\href@noop {} {\bibfield  {journal} {\bibinfo  {journal}
  {Phys. Rev. E}\ }\textbf {\bibinfo {volume} {57}},\ \bibinfo {pages} {4817}
  (\bibinfo {year} {1998})}\BibitemShut {NoStop}%
\bibitem [{\citenamefont {Kunihiro}\ and\ \citenamefont
  {Tsumura}(2006)}]{kunihiro2006application}%
  \BibitemOpen
  \bibfield  {author} {\bibinfo {author} {\bibfnamefont {T.}~\bibnamefont
  {Kunihiro}}\ and\ \bibinfo {author} {\bibfnamefont {K.}~\bibnamefont
  {Tsumura}},\ }\href@noop {} {\bibfield  {journal} {\bibinfo  {journal} {J.
  Phys. A: Math. Gen.}\ }\textbf {\bibinfo {volume} {39}},\ \bibinfo {pages}
  {8089} (\bibinfo {year} {2006})}\BibitemShut {NoStop}%
\bibitem [{\citenamefont {Chiba}\ and\ \citenamefont
  {Hayato}(2008)}]{Chiba2008Approximation}%
  \BibitemOpen
  \bibfield  {author} {\bibinfo {author} {\bibnamefont {Chiba}}\ and\ \bibinfo
  {author} {\bibnamefont {Hayato}},\ }\href@noop {} {\bibfield  {journal}
  {\bibinfo  {journal} {J. Math. Phys.}\ }\textbf {\bibinfo {volume} {49}},\
  \bibinfo {pages} {102703} (\bibinfo {year} {2008})}\BibitemShut {NoStop}%
\bibitem [{\citenamefont {Watts}\ and\ \citenamefont
  {Strogatz}(1998)}]{watts1998collective}%
  \BibitemOpen
  \bibfield  {author} {\bibinfo {author} {\bibfnamefont {D.~J.}\ \bibnamefont
  {Watts}}\ and\ \bibinfo {author} {\bibfnamefont {S.~H.}\ \bibnamefont
  {Strogatz}},\ }\href@noop {} {\bibfield  {journal} {\bibinfo  {journal}
  {Nature}\ }\textbf {\bibinfo {volume} {393}},\ \bibinfo {pages} {440}
  (\bibinfo {year} {1998})}\BibitemShut {NoStop}%
\bibitem [{\citenamefont {Marconi}(1909)}]{marconi1909wireless}%
  \BibitemOpen
  \bibfield  {author} {\bibinfo {author} {\bibfnamefont {G.}~\bibnamefont
  {Marconi}},\ }\href@noop {} {\bibfield  {journal} {\bibinfo  {journal} {Nobel
  Lecture}\ }\textbf {\bibinfo {volume} {11}},\ \bibinfo {pages} {198}
  (\bibinfo {year} {1909})}\BibitemShut {NoStop}%
\end{thebibliography}

%

\end{document}